\documentclass[10pt,amsmath,noshowkeys,noshowpacs,longbibliography,amssymb,aps,pra,floatfix]{revtex4-2}

\usepackage[utf8]{inputenc}
\usepackage[T1]{fontenc}
\usepackage{graphicx}
\usepackage{geometry}
\usepackage{xcolor}
\usepackage{amsmath,amssymb}
\usepackage{bm}
\usepackage[hidelinks]{hyperref}
\usepackage{booktabs}
\usepackage{float}
\usepackage{subcaption}

\begin{document}

\title{Approximating velocity fields with planted point attractors via Neural-ODEs for classification purposes}

\author{Feliciano Giuseppe Pacifico $^{1,2}$, Duccio Fanelli $^{2}$, Lorenzo Buffoni $^2$, Lorenzo Chicchi $^{2}$, Diego Febbe $^2$, Raffaele Marino $^{2}$}

\affiliation{$1$ Department of Informatics and Computer Science, University of Pisa, Italy }
\affiliation{$2$ Department of Physics and Astronomy, University of Florence, Sesto Fiorentino, Italy \\ INFN, Italy}

 \begin{abstract}
In this work, Neural-ODEs equipped with a curated collection of equilibrium points have been successfully employed for classification tasks. The planted fixed points serve as indicators for the target classes, while the velocity field, leveraging the universal approximation capabilities of the architecture, shapes the dynamical landscape. This process defines the basins of attraction of the trained model, effectively directing each input (provided as an initial condition) toward its corresponding destination target.
\end{abstract}

\maketitle
\section{Introduction}

Neural ordinary differential equations (Neural-ODEs)~\cite{chen2018neuralode} define a 
class of deep learning models designed to parameterize the derivative of a multi-dimensional state vector, via a neural network. In practice, the forward pass is framed as the trajectory of a point within a high-dimensional manifold, governed by a system of coupled ODEs. Here, a deep neural network, leveraging its universal approximation properties, defines the underlying vector field \cite{teshima2020universal, ishikawa2023universal}.
Neural-ODEs serve as effective feature extractors: inputs are treated as initial conditions for a learnable dynamical system, with the resulting terminal state being classified by an external readout, such as a linear classifier or a compact Multi-Layer Perceptron.

A complementary line of work has recently been proposed \cite{GaglianiPacificoChicchiFanelliFebbeBuffoniMarino2026} that aims at performing classification by encoding the decision rule in the phase portrait of a dynamical model. Each class to be eventually identified
is represented by a stable point attractor (also referred to as simply attractor in the following): the label is the attractor reached by the flow. Following this approach, the dynamical system itself is the classifier: no additional readout head is needed to perform the task. Different dynamical models can be adapted to this purpose by devising suitable strategies to plant stationary attractors, i.e.\ fixed points that are imposed by design and subsequently used as class prototypes. Applications range from the SA-Neural-ODE setting \cite{Marino_2024}, where each individual component is subject to a bistable potential and the nodes that define the computing network are linearly entangled, to Hopfield-like models, where non-local interactions are filtered by a sigmoidal function \cite{Chicchi_2025, GaglianiPacificoChicchiFanelliFebbeBuffoniMarino2026}, and to the Wilson-Cowan model \cite{wilson1972cowan,marino2025learning}, a standard multi-population reference for computational neuroscience modeling. In all these cases the fixed points are planted ex ante by constraining a limited subset of the model's parameters, for instance through a spectral construction of the coupling matrix as in \cite{Chicchi_2025, marino2025learning}; the remaining parameters are trained to generate a velocity field that steers the initial conditions, the items to be classified, toward the designated equilibrium, thus flagging the class they belong to. Upon training the fixed points are thus turned into veritable point attractors, stability being enforced as a byproduct of the training. A key theoretical gap, however, remains between the planted-attractor models just listed and the Neural-ODE framework. Planted-attractor constructions enhance interpretability, as learning implies enforcing stability and sculpting the associated basins of attraction of the target equilibria within a dynamical framework: the trained model can then be analyzed with the well-established tools of dynamical systems theory, and is in this sense mechanistically more interpretable. On the other hand, the resulting architecture, tied to a specific closed-form interaction term and constrained to accommodate the imposed attractors, is not covered by the standard universal approximation theorems: it might prove unable to approximate the velocity field required by a given task, which would then be out of reach by design. In the Neural-ODE scheme the opposite holds: universality is guaranteed, but classification requires additional processing, such as an MLP head acting on the terminal state, as no attractors are enforced a priori in the velocity field. The aim of this work is, in the first place, to unify the two views: we let a Neural-ODE perform classification by shaping the phase space of the dynamical system, so as to obtain a classifier that (i) retains universality and (ii) can at the same time be studied mechanistically with the tools of physics, dynamical systems theory and statistical mechanics, and is in this sense more interpretable.

In \cite{pacificoNEURIPS}, a technique was introduced to incorporate a finite set of fixed points, i.e.\ points where the velocity field is identically zero, within the multi-dimensional state space of a Neural-ODE. This approach ensures that gradient-based training remains strictly constrained within the prescribed hypothesis class without compromising the model's expressive power. The universality of this modified Neural-ODE framework was established there, together with a computationally efficient methodology for enforcing the forced equilibria.

To consolidate the concept of dynamical classifier, we here leverage the generalized version of Neural-ODEs recalled above. By capitalizing on forced equilibria to represent specific classes, we establish a novel classification framework that enjoys a proven universal approximation property. This distinguishes our methodology from earlier work \cite{GaglianiPacificoChicchiFanelliFebbeBuffoniMarino2026}, which sought to harmonize classification with dynamical properties using different constraints. We stress from the outset that the resulting classifier does not improve upon a standard MLP with the same number of parameters in terms of accuracy, training cost or forward FLOPs: on the benchmarks considered here it reaches a comparable accuracy, at a larger computational cost. What it provides is an alternative approach to classification, whose mechanism can be interpreted in dynamical terms, in the sense made precise in Sec.~\ref{subsec:interpretability}. Beyond the mechanistic interpretability that comes with the use of a dynamical system, a further step can be taken. Building on the scheme of Funahashi and Nakamura \cite{FUNAHASHI1993801}, the Neural-ODE with planted fixed points can be read as the singular limit of a continuous-time recurrent network with a visible and a hidden population, which interact and evolve on distinct time scales. Models of interacting populations are well known in the theory of dynamical systems, and it is therefore of interest to look at the classifier in these terms as well. The universality that comes with the modified Neural-ODEs is inherited, in the same limit, by the two-population scheme.

The paper is organized as follows. In the next section we recall the recipe to yield generalized Neural-ODEs with planted fixed points, and we introduce their two-population analogue. The subsequent section is devoted to the modified Neural-ODEs operated in
classification mode, extending beyond the settings explored so far in the literature. Finally, we sum up and draw our conclusions.

\section{Planting fixed points within a Neural-ODE while preserving universality} \label{sec:bridge_ch1_ch2}

Consider a Neural-ODE model of the form:

\begin{equation}
\dot x(t) = F_\theta(x(t)), 
\qquad
F_\theta(x) := -x + A_1 f(A_2 x + b_2) + b_1,
\label{eq:node_model}
\end{equation}

where $x(t)\in\mathbb R^n$, $A_2\in\mathbb R^{m\times n}$, $b_2\in\mathbb R^m$,
$A_1\in\mathbb R^{n\times m}$, $b_1\in\mathbb R^n$, and $f:\mathbb R^m\to\mathbb R^m$
is applied component-wise (e.g.\ $f(z)_j=\sigma(z_j)$). The right-hand side of (\ref{eq:node_model}) contains a linear term ($-x$). This choice is motivated by two primary factors. First, the term introduces a stabilizing influence, which is beneficial for classification: it favors the relaxation of the trajectories towards the equilibria. Second, the model in (\ref{eq:node_model}) can be mapped onto an equivalent scheme of multi-species binary interactions governed by the weighted adjacency matrices $A_1$ and $A_2$, as discussed in Sec.~\ref{subsec:epsilon_nonzero} \cite{FUNAHASHI1993801}. Finally, for completeness, we include biases $b_1$ and $b_2$ in both the hidden and output layers.

In \cite{pacificoNEURIPS}, we have (i) introduced an algorithmic procedure 
to insert a set of $C$ fixed points in the above vector field $F_\theta(x)$, namely $F_\theta(\bar x^{(\ell)}) = 0,\qquad \ell=1,\dots,C$; (ii) proven that imposing hard equilibrium constraints does not reduce the expressive power of the Neural-ODE vector field: on compact sets, the constrained model class remains dense in the class of continuous velocity fields that vanish at the prescribed points. In Sec.~\ref{subsec:node_setup} we briefly recall only the planting procedure (i), which is used throughout the paper, before exploring the potential of the Neural-ODE model, complemented with a set of prescribed fixed points, in classification mode. For the derivation of the procedure in full detail, and for the formal statement and proof of the universality result (ii), the reader is referred to \cite{pacificoNEURIPS}. Note that enforcing a priori a finite set of prescribed fixed points within the adjustable velocity field offers a distinct algorithmic advantage: the constraints ensuring the existence of these fixed points hold for all parameter values. Consequently, gradient-based training is structurally prevented from drifting outside the constrained hypothesis class. The planted fixed points are thus preserved throughout optimization without the need for auxiliary loss penalties or post-hoc corrections. The QR-planted Neural-ODEs will be hereafter referred to as ConsNODEs, for Constrained Neural-ODEs.

\subsection{Imposing the fixed-point constraints}
\label{subsec:node_setup}

Given a set of $C$ desired equilibria $\{\bar{x}^{(\ell)}\}_{\ell=1}^C \subset \mathbb{R}^n$, which serve as class prototypes when the model operates in classification mode, the fixed-point constraints are defined as follows:

\begin{equation}
F_\theta(\bar x^{(\ell)})=0,
\qquad \ell=1,\dots,C.
\label{eq:fp_constraints}
\end{equation}

Evaluating \eqref{eq:node_model} at $\bar x^{(\ell)}$ yields:
\[
0
=
-\bar x^{(\ell)} + A_1 f(A_2 \bar x^{(\ell)} + b_2) + b_1
\quad\Longleftrightarrow\quad
A_1 s_\ell = y_\ell,
\]
where the \emph{feature vectors} and \emph{targets} respectively read:
\begin{equation}
s_\ell := f(A_2 \bar x^{(\ell)} + b_2)\in\mathbb R^m,
\qquad
y_\ell := \bar x^{(\ell)} - b_1\in\mathbb R^n.
\label{eq:def_s_y}
\end{equation}
Next we stack the above quantities into matrices
\begin{equation}
S := [s_1,\dots,s_C]\in\mathbb R^{m\times C},
\qquad
Y := [y_1,\dots,y_C]\in\mathbb R^{n\times C}.
\label{eq:def_S_Y}
\end{equation}
Consequently, the $C$ fixed-point constraints \eqref{eq:fp_constraints} can be condensed into a single matrix equation:
\begin{equation}
A_1 S = Y.
\label{eq:A1S_eq_Y}
\end{equation}
Crucially, for any fixed set of parameters $(A_2, b_2, b_1)$, the constraint \eqref{eq:A1S_eq_Y} remains \emph{linear} with respect to $A_1$.

Assuming that the planted features are linearly independent (namely, that $\mathrm{rank}(S)=C$ with $m > C$), the linear system \eqref{eq:A1S_eq_Y} is guaranteed to be consistent and admits an infinite number of solutions. To characterize this solution space, we consider the thin QR decomposition \cite{GolubVanLoan2013, BenIsraelGreville2003} of the feature matrix:\begin{equation}S = QR,\label{eq:qr_S}\end{equation}where $Q \in \mathbb{R}^{m \times C}$ possesses orthonormal columns ($Q^\top Q = I_C$) and $R \in \mathbb{R}^{C \times C}$ is an invertible upper triangular matrix. 

The assumption $\mathrm{rank}(S)=C$ involves the planted equilibria, since $S$ depends on them, but it restricts their choice only in that they must be pairwise distinct: if two of them coincide, two columns of $S$ are identical for every value of the parameters. For pairwise distinct equilibria, and for any continuous activation $f$ that is not a polynomial, there exist instead parameters $(A_2,b_2)$ for which $S$ has full column rank \cite{Pinkus1999}, and such parameters are not exceptional. The matrices of rank smaller than $C$ form a closed set of zero measure in $\mathbb R^{m\times C}$, so that the matrices of full rank are an open and dense set; correspondingly, for an analytic activation such as the sigmoid, $\det(S^\top S)$ is an analytic function of $(A_2,b_2)$ that is not identically zero, and it vanishes only on a closed set of zero Lebesgue measure. The equilibria can therefore be placed freely: an initialization of $A_2$ and $b_2$ drawn from a continuous distribution yields a matrix $S$ of full rank with probability one (for the ReLU, with a probability that rapidly approaches one, as $m$ grows beyond $C$), and in practice the rank of $S$ is verified at initialization. The position of the equilibria affects instead the conditioning of $S$, as discussed in Appendix~\ref{sec:attractor_distance}.

The procedure outlined in \cite{pacificoNEURIPS} consists of the following steps:

\begin{enumerate}
    \item Projector via QR. Given that $\mathrm{span}(S) = \mathrm{span}(Q)$, the orthogonal projector onto the subspace spanned by the features is defined as:
\begin{equation}
P = QQ^\top.
\label{eq:P_QQt}
\end{equation}
It follows immediately that $(I - P)S = 0$, as each column of $S$ by definition lies within the range of $Q$. This identity ensures that the null space of the operator $(I - P)$ completely contains the set of planted features.

\item To satisfy the constraint $A_{\mathrm{part}}S=Y$, we substitute the QR decomposition $S=QR$ to obtain:$$A_{\mathrm{part}}QR = Y.$$By defining the intermediate matrix $B := A_{\mathrm{part}}Q \in \mathbb{R}^{n \times C}$, the system simplifies to:\begin{equation}BR = Y.\label{eq:BR_eq_Y}\end{equation}Since $R$ is an invertible upper triangular matrix, \eqref{eq:BR_eq_Y} admits the unique solution $B = YR^{-1}$, yielding the particular solution:\begin{equation}A_{\mathrm{part}} = BQ^\top = (YR^{-1})Q^\top.\label{eq:A_part_qr}\end{equation}From a computational standpoint, explicitly computing $R^{-1}$ is unnecessary. In practice, the system $BR=Y$ is solved via back-substitution, a numerically stable operation with a complexity of $O(C^2)$ per output dimension.
\item Analogously to the pseudoinverse formulation, the full family of solutions is obtained by appending a term that lies in the null space of $S$:
\begin{equation}
A_1 := A_{\mathrm{part}} + W(I-P)=(YR^{-1}Q^\top) + W(I-QQ^\top),\qquad W\in\mathbb R^{n\times m},\label{eq:A1_qr_param}
\end{equation}
where $W$ represents a matrix of free, tunable parameters. It follows that:$$A_1 S
=
A_{\mathrm{part}}S + W(I-P)S
=
Y + W \cdot 0
=
Y.$$Consequently, the fixed-point constraints \eqref{eq:fp_constraints} are satisfied exactly, regardless of the values assigned to $W$. This parametrization ensures that the existence of the planted fixed points is maintained throughout any gradient-based optimization of the weight matrix $W$.
\end{enumerate}

\subsection{Stability of the planted equilibria: closed form of the Jacobian}
\label{sec:jacobian_stability}

The planting procedure of Sec.~\ref{sec:bridge_ch1_ch2} guarantees that the class prototypes $\bar x^{(\ell)}$ are equilibria of the learned flow, but it does not by itself make them \emph{attractive}. Local stability can be probed by a linearization of $F_\theta$ around any planted equilibrium $\bar x^{(\ell)}$. In this section we carry out this linearization in closed form. In the numerical experiments reported below, the spectrum of the Jacobian matrix will be analyzed for the different models examined. We will in particular show that even when the equilibria are deliberately initialized so that \emph{every} direction is unstable, training on the classification loss alone drives them to become stable.

Let $z(x):=A_2x+b_2\in\mathbb R^m$ be the pre-activation of the hidden layer, so that $F_\theta(x)=-x+A_1f(z(x))+b_1$. Since $f$ acts component-wise, $\partial f(z)/\partial z=\operatorname{diag}\big(f'(z_1),\dots,f'(z_m)\big)$, and the chain rule gives
\begin{equation}
J_\theta(x):=\frac{\partial F_\theta}{\partial x}(x)
=-I_n+A_1\,D(x)\,A_2,
\qquad
D(x):=\operatorname{diag}\!\big(f'(z_1(x)),\dots,f'(z_m(x))\big)\in\mathbb R^{m\times m}.
\label{eq:jacobian_general}
\end{equation}
At a planted equilibrium the diagonal matrix is evaluated at the same pre-activations that define the feature vector $s_\ell=f(A_2\bar x^{(\ell)}+b_2)$ of Eq.~\eqref{eq:def_s_y},
\begin{equation}
J^{(\ell)}:=J_\theta\big(\bar x^{(\ell)}\big)=-I_n+A_1D_\ell A_2,
\qquad D_\ell:=D\big(\bar x^{(\ell)}\big).
\label{eq:jacobian_planted}
\end{equation}
Inserting the parametrization \eqref{eq:A1_qr_param} of $A_1$ we obtain the closed form in terms of the trainable quantities,
\begin{equation}
J^{(\ell)}=-I_n+\Big[\,YR^{-1}Q^\top+W\,(I_m-QQ^\top)\Big]\,D_\ell\,A_2 ,
\label{eq:jacobian_param}
\end{equation}
where $Y$, $Q$ and $R$ depend on $(A_2,b_2,b_1)$ through the planted features $S$ (Eqs.~\eqref{eq:def_S_Y} and \eqref{eq:qr_S}). For the ReLU activation used in the Fashion-MNIST experiments (see below), $f'(z)=\Theta(z)$ is the Heaviside function: $D_\ell$ is then a diagonal $0/1$ matrix that selects the hidden units active at $\bar x^{(\ell)}$, and \eqref{eq:jacobian_planted} is exact.

Two remarks are in order. First, the planting fixes the first term in the square bracket of \eqref{eq:jacobian_param}, but the free matrix $W$ enters $J^{(\ell)}$ through $W(I_m-QQ^\top)D_\ell A_2$. Since $W(I_m-QQ^\top)S=0$, the matrix $W$ leaves the existence of the equilibria untouched, and yet it modifies their linear stability. Stability is therefore not granted by the construction: it is a property that has to be selected by the training. Second, by Lyapunov's indirect method $\bar x^{(\ell)}$ is asymptotically stable if all the eigenvalues of $J^{(\ell)}$ have negative real part, i.e.\ if the spectral abscissa $\alpha_\ell:=\max_i\operatorname{Re}\lambda_i(J^{(\ell)})$ is negative, and it is unstable if $\alpha_\ell>0$. Writing $M_\ell:=A_1D_\ell A_2$, the leak term shifts the spectrum rigidly, $\lambda_i(J^{(\ell)})=\mu_i(M_\ell)-1$, so that the term $-x$ stabilizes the equilibrium only as long as $\operatorname{Re}\mu_i(M_\ell)<1$ for every $i$.

\subsection{Interacting populations: a multi-species variant of the Neural-ODE with planted fixed points}
\label{subsec:epsilon_nonzero}

In this subsection we discuss a closely related \emph{coupled} dynamical system, a continuous-time recurrent network with a visible population $x$ and a hidden population $u$, that reduces to the planted Neural-ODE model \eqref{eq:node_model} in the singular limit $\varepsilon\to 0$, where $\varepsilon$ stands for an appropriate time scale, as introduced below. This viewpoint---which builds on the Funahashi--Nakamura scheme \cite{FUNAHASHI1993801}---is useful both conceptually (it interprets the nonlinearity as a fast auxiliary subsystem) and practically (it provides an augmented-state realization whose trajectories, when projected onto $x$,
match those of the planted Neural-ODE for sufficiently small $\varepsilon$).

Starting from the parameters $(A_1,A_2,b_1,b_2)$ of the planted Neural-ODE \eqref{eq:node_model},
consider the following augmented system in the variables $(x(t),u(t))\in\mathbb R^n\times\mathbb R^m$:
\begin{equation}
\label{eq:coupled_model}
\begin{cases}
\dot x(t) = -x(t) + A_1 f\!\big(u(t)\big) + b_1,\\[4pt]
\varepsilon\,\dot u(t) = -u(t) + A_2 x(t) + b_2,
\end{cases}
\qquad \varepsilon>0.
\end{equation}
This model will be referred to as the 2PoP model in the following.
For small $\varepsilon$, $u(t)$ evolves on a fast time scale and acts as a rapidly relaxing proxy for the
pre-activation $A_2x(t)+b_2$. Notice that \eqref{eq:coupled_model} is an \emph{autonomous} ODE in the extended
state $(x,u)$, while the projected dynamics in $x$ alone is generally \emph{not} described by a single-valued
vector field $F(x)$ because $\dot x$ depends on the additional variable $u(t)$.

In the limit $\varepsilon\to 0$ one can adiabatically eliminate the variable $u$. This amounts to setting the left-hand side of the second of Eqs.~(\ref{eq:coupled_model}) to zero, which yields

\begin{equation}
\label{eq:slow_manifold_constraint}
u(t)=u^{*}(x(t))=A_2x(t)+b_2.
\end{equation}
Substituting \eqref{eq:slow_manifold_constraint} into the $x$-equation gives the reduced dynamics
\begin{equation}
\label{eq:reduced_model_equals_model1}
\dot x(t)
=
-x(t) + A_1 f\!\big(A_2x(t)+b_2\big) + b_1,
\end{equation}
which coincides exactly with the planted Neural-ODE vector field \eqref{eq:node_model}.
In other words, the planted Neural-ODE can be interpreted as the $\varepsilon\to 0$ quasi-steady reduction of the coupled
system \eqref{eq:coupled_model}.

For $\varepsilon>0$ the set
\begin{equation}
\label{eq:slow_manifold_def}
\mathcal M := \{(x,u)\in\mathbb R^n\times\mathbb R^m:\ u=u^{*}(x)=A_2x+b_2\}
\end{equation}
is an invariant manifold of the singular limit $\varepsilon\to 0$ and is attracting because the Jacobian of the fast dynamics (derivative 
with respect to $u$) is $-\varepsilon^{-1}I_m$.
Standard fast--slow arguments \cite{Kristiansen2023, SerinoAlvarezLoyaBurbyKevrekidisTang2025} therefore imply that, after a short transient of duration $O(\varepsilon)$,
trajectories $(x_\varepsilon(t),u_\varepsilon(t))$ of \eqref{eq:coupled_model} remain close to $\mathcal M$ and the
projected evolution $x_\varepsilon(t)$ tracks the reduced flow \eqref{eq:reduced_model_equals_model1} on finite horizons.
Both distances are of order $\varepsilon$. Writing $e:=u-(A_2x+b_2)$, the second of Eqs.~\eqref{eq:coupled_model} gives $\varepsilon\dot e=-e-\varepsilon A_2\dot x$, so that after the transient $e\simeq-\varepsilon A_2F_\theta(x)$ to first order in $\varepsilon$: the trajectories do not lie on $\mathcal M$, but on a nearby invariant manifold at a distance $O(\varepsilon)$ from it, and $x_\varepsilon(t)-x(t)=O(\varepsilon)$ on finite horizons, $x(t)$ being the solution of \eqref{eq:reduced_model_equals_model1} with the same initial condition.
This explains why, when $\varepsilon$ is sufficiently small, the qualitative phase portrait observed in the $x$-space
matches that of the planted Neural-ODE even though \eqref{eq:coupled_model} is defined in an extended state space. A numerical illustration of the convergence for a specific dataset is given in Sec.~\ref{subsec:two_pop_spirals}.

\subsection{Interpretability: dynamical and population-level perspectives}
\label{subsec:interpretability}

Interpretability is a multifaceted concept in machine learning, broadly associated with the possibility of understanding how a model operates and produces its predictions. Although the terms \emph{interpretability} and \emph{explainability} are often used interchangeably in the literature, a useful conceptual distinction can be drawn between understanding the internal mechanisms of a model and providing explanations of its predictions \cite{lipton2018mythos, doshivelez2017towards, rudin2019stop}.

To illustrate this distinction, consider a neural network trained to classify images. A post-hoc explainability method, such as a saliency map \cite{simonyan2014deep} or feature-attribution technique \cite{lundberg2017unified, chicchi2026estimating}, may identify which pixels or image regions contributed most strongly to a particular prediction. Such an analysis can help explain why a given input was assigned to a certain class, without necessarily revealing the internal computational mechanism responsible for that decision. By contrast, a simple decision tree can be considered intrinsically interpretable because its classification rules can be directly inspected and followed from input to output. More generally, a mechanistic perspective on interpretability seeks to understand the internal processes through which a trained model implements a given computation, rather than restricting the analysis to relations between input features and predictions. These different notions are not mutually exclusive, nor do they establish a universal distinction between interpretability and explainability. They nevertheless provide a useful framework for specifying what kind of understanding a model makes accessible.

In the present work, we focus on a particular form of mechanistic understanding, which we refer to as \emph{dynamical interpretability}. Once trained, the proposed Neural-ODE defines an explicitly specified autonomous vector field on a fixed state space. Unlike generic feedforward architectures, where successive layers may transform representations across spaces of different dimensions, the evolution of the Neural-ODE takes place within the same state space and is governed by a single vector field. In mathematical terms, the learned vector field maps the state space into itself. The resulting flow therefore provides a coherent geometrical description of the computational path performed by the model.

This viewpoint allows one to investigate the classification mechanism at the level of the learned dynamics, without necessarily reconstructing the contribution of every individual neuron or parameter used to represent the vector field. In particular, one can study the evolution of positions and velocities, the geometry of trajectories and of the learned flow, the location and local stability of fixed points through the Jacobian spectrum, convergence properties, and the structure of the corresponding basins of attraction. Such analyses may reveal how the trained system transports different inputs, how competing class assignments evolve along trajectories, and how the dynamical landscape supports the final classification. The use of fixed-point and linear stability analyses to uncover computational mechanisms in trained recurrent neural networks has already been explored in the literature \cite{sussillo2013opening}. Such an analysis can also reveal what the accuracy hides: in the example of Sec.~\ref{subsec:trap_disk_2d}, a model with a high accuracy turns out to send one of the classes not to its planted equilibrium, but to an attractor that the training has created.

In our framework, this dynamical perspective is further strengthened by the presence of prescribed class equilibria. Their positions are known by construction and remain fixed throughout training, providing explicit reference states around which the learned dynamics can be investigated. Although the planting procedure does not automatically guarantee their stability, the training process can shape the surrounding vector field so that these equilibria become attractive. Consequently, the classification process can be studied in terms of the evolution toward prescribed states, the stability properties acquired during training, and the geometry of the corresponding attraction regions. This does not imply that the resulting nonlinear system is automatically transparent, nor that every aspect of its behavior can be characterized analytically. Rather, it provides a structured mathematical framework in which the mechanisms underlying classification can be investigated quantitatively using established tools from dynamical-systems theory.

A second, complementary perspective follows from the interacting-population formulation introduced above (Sec.~\ref{subsec:epsilon_nonzero}). In the spirit of Funahashi and Nakamura \cite{FUNAHASHI1993801}, the Neural-ODE can be recovered as the singular limit of a continuous-time recurrent system comprising visible and hidden populations evolving on different time scales. The matrices governing the neural-network transformations can then be interpreted as weighted interaction matrices between these populations. More precisely, $A_2$ carries the signal from the visible population to the hidden one, and $A_1$ from the hidden population back to the visible one. This reading enters the stability analysis of Sec.~\ref{sec:jacobian_stability}: in the limit $\varepsilon\to0$, a perturbation $\delta x$ of the visible population around the planted equilibrium $\bar x^{(\ell)}$ returns to it, through the hidden population, as $A_1D_\ell A_2\,\delta x$, where $D_\ell$ weights each hidden unit by the slope of its activation at the equilibrium. The matrix $M_\ell=A_1D_\ell A_2$ that appears in the Jacobian \eqref{eq:jacobian_planted} is thus the gain of the loop between the two populations, and the condition $\operatorname{Re}\mu_i(M_\ell)<1$ states that the equilibrium is stable as long as this gain does not overcome the leak term $-x$. This representation connects the proposed classifier with the broader tradition of population-based neural dynamics studied in theoretical neuroscience, including Wilson--Cowan and continuous-time Hopfield-like models \cite{wilson1972cowan, hopfield1984graded}. The analogy concerns the form of the equations, and not the existence of an energy function: at variance with the Hopfield model, the couplings $A_1$ and $A_2$ are not constrained by any symmetry, so that the existence of a Lyapunov function is not guaranteed. Convergence towards the planted equilibria and the structure of their basins are therefore not inherited from the Hopfield theory: they are properties of the trained model, to be assessed with the dynamical analysis described above.

Although this correspondence does not imply that the trained system constitutes a biologically faithful model of neuronal activity, it offers an additional structural interpretation of its computational organization. The classification task, performed in the visible state space, can be viewed as emerging from the interactions between the visible and auxiliary hidden populations. This perspective also opens the possibility of investigating collective dynamical properties, interaction structures, and statistical-mechanical characteristics of the trained classifier, which we leave for future work.

\section{Classification via constrained Neural-ODEs}
\label{sec:classification}

We now turn to applying the modified Neural-ODE with planted fixed points to a wide range of classification tasks. The purpose of this section is twofold: (i) to present, at a conceptual level, how classification can be
performed by a dynamical system with \emph{hard-planted} class equilibria, and (ii) to quantify its performance and computational cost on benchmark datasets (here we deal with Fashion-MNIST \cite{xiao2017fashion} and CIFAR-10 \cite{Krizhevsky2009}) as compared to a parameter-matched static Multi-Layer Perceptron (MLP) head.

\subsection{Planted-attractor dynamical classification: general principle}
\label{subsec:dynamical_classification_principle}
Let $C$ be the number of classes. We prescribe a set of $C$ distinct target equilibria
$\{\bar x^{(\ell)}\}_{\ell=1}^C \subset \mathbb R^n$, one for each class.
Given an input $u$ (e.g., an image), an encoder $\phi$ (the identity, if no pre-processing is needed) maps it to an initial condition in state space,
\begin{equation}
x(0)=x_0=\phi(u)\in\mathbb R^n.
\end{equation}
We then evolve $x(t)$ under a learned autonomous vector field $F_\theta:\mathbb R^n\to\mathbb R^n$,
\begin{equation}
\dot x(t)=F_\theta(x(t)),\qquad x(0)=x_0,
\label{eq:generic_node}
\end{equation}
up to a fixed terminal time $T>0$, yielding the terminal state $x(T)=\Psi_T(x_0)$, where $\Psi_T$ denotes
the flow map at time $T$. We enforce the equilibrium constraints
\begin{equation}
F_\theta(\bar x^{(\ell)})=0,\qquad \ell=1,\dots,C,
\label{eq:equilibrium_constraints}
\end{equation}
\emph{by construction} via the QR-based projector parametrization discussed above.
Recalling the above discussion, we rewrite \eqref{eq:equilibrium_constraints} as a linear system $A_1S=Y$, compute a thin QR factorization $S=QR$ and set
\begin{equation}
A_1=(YR^{-1})Q^\top + W(I-QQ^\top),
\label{eq:cifar_A1_qr}
\end{equation}
where $W$ and $A_2,b_2,b_1$ are free and trainable. Since \eqref{eq:cifar_A1_qr} implies
$A_1S=Y$, all planted equilibria are preserved exactly for \emph{all} values of the free parameters; in particular,
gradient-based training cannot drift away from the constrained manifold.

For a labeled sample $(u,\ell)$ with $\ell\in\{1,\dots,C\}$, we train the Neural-ODE model, supplemented with the required fixed points, so that the terminal state
$x(T;u)$ is close to the corresponding class equilibrium $\bar x^{(\ell)}$. Concretely, we minimize a
terminal-time regression loss
\begin{equation}
\mathcal L_{\mathrm{task}}(\theta)
=\mathbb E_{(u,\ell)}\Big[\big\|x(T;u)-\bar x^{(\ell)}\big\|_2^2\Big],
\label{eq:terminal_regression_loss}
\end{equation}
where the expectation is approximated by the empirical average over mini-batches.

At test time, classification is performed by selecting the planted equilibrium that is nearest to the terminal state:
\begin{equation}
\widehat y \;:=\; \arg\min_{\ell\in\{1,\dots,C\}} \bigl\|x(T)-\bar x^{(\ell)}\bigr\|_2^2.
\label{eq:nearest_attractor_rule}
\end{equation}
Intuitively, training shapes the vector field so that encoded samples $x_0=\phi(u)$ are driven, within time $T$,
towards the equilibrium associated with their label. The central structural constraint in our approach is that the
targets $\bar x^{(\ell)}$ are \emph{exact} equilibria of the learned dynamics for all parameter values encountered
during optimization. Moreover, the Neural-ODE with planted equilibria is a universal approximator of the velocity fields compatible with the planted equilibria. This provides the system with the needed flexibility for solving virtually any classification problem, with one caveat. The flow map $\Psi_T$ is the flow of an autonomous ODE, hence a homeomorphism of $\mathbb R^n$ onto its image, and inherits its topology from the Cauchy theorem; trajectories of \eqref{eq:generic_node} cannot cross. If a class is nested inside another, e.g.\ a disk surrounded by an annulus of the other class in $\mathbb R^2$, the encircled class cannot be driven to its equilibrium without crossing the trajectories of the class that surrounds it, and training instead produces a spurious internal attractor. Enlarging the embedding dimension $n$ removes the obstruction, since the two classes can then pass each other through the added dimensions; we shall make this statement precise and demonstrate it on the disk-and-annulus example reported in Sec.~\ref{sec:topological_trap}.

In our experiments we use a simple block-coded family of equilibria. The coordinates of $\mathbb R^n$ are partitioned
into $C$ disjoint blocks of equal size $n/C$, and $\bar x^{(\ell)}$ is defined by setting a fixed amplitude $a>0$ on
block $\ell$ and $0$ elsewhere. Any fixed collection of distinct class equilibria would be suitable. What matters is
that the same set $\{\bar x^{(\ell)}\}$ is used consistently for training and decoding.

\subsection{Toy experiment: two interleaving spirals}
\label{subsec:spirals_toy}

Before moving to high-dimensional image representations, we first illustrate the planted-at\-trac\-tor
dynamical-clas\-si\-fi\-ca\-tion mechanism on a standard nonlinearly separable toy dataset: the
\emph{two interleaving spirals} in $\mathbb R^2$.
This setting is particularly informative because the state space is two-dimensional, hence the
learned phase portrait, basins of attraction, and induced decision regions can be visualized directly.

We generate two planar spirals with one full turn, and construct a balanced dataset of
$N=30{,}000$ labeled points $x_0\in\mathbb R^2$ split into train/validation/test with ratios $0.6/0.2/0.2$.
In this experiment the encoder is the identity map, i.e.\ the input point is used directly as initial condition
$x(0)=x_0$.

We consider $C=2$ classes and plant two target equilibria in $\mathbb R^2$,
\[
\bar x^{(1)}=\begin{pmatrix} 1\\ 0\end{pmatrix},
\qquad
\bar x^{(2)}=\begin{pmatrix}-1\\ 0\end{pmatrix}.
\]
We instantiate the ConsNODEs scheme, i.e.\ the QR-planted Neural-ODE head, with state dimension $n=2$ and hidden width $m=300$,
using a sigmoid nonlinearity. The dynamics are integrated with an explicit Euler discretization with step size
$\Delta t=0.03$ for $40$ steps (terminal time $T=1.2$).
Training minimizes the terminal regression loss \eqref{eq:terminal_regression_loss}, and prediction uses the
nearest-equilibrium rule \eqref{eq:nearest_attractor_rule}.

\begin{figure}[H]
    \centering
    \includegraphics[width=0.98\linewidth]{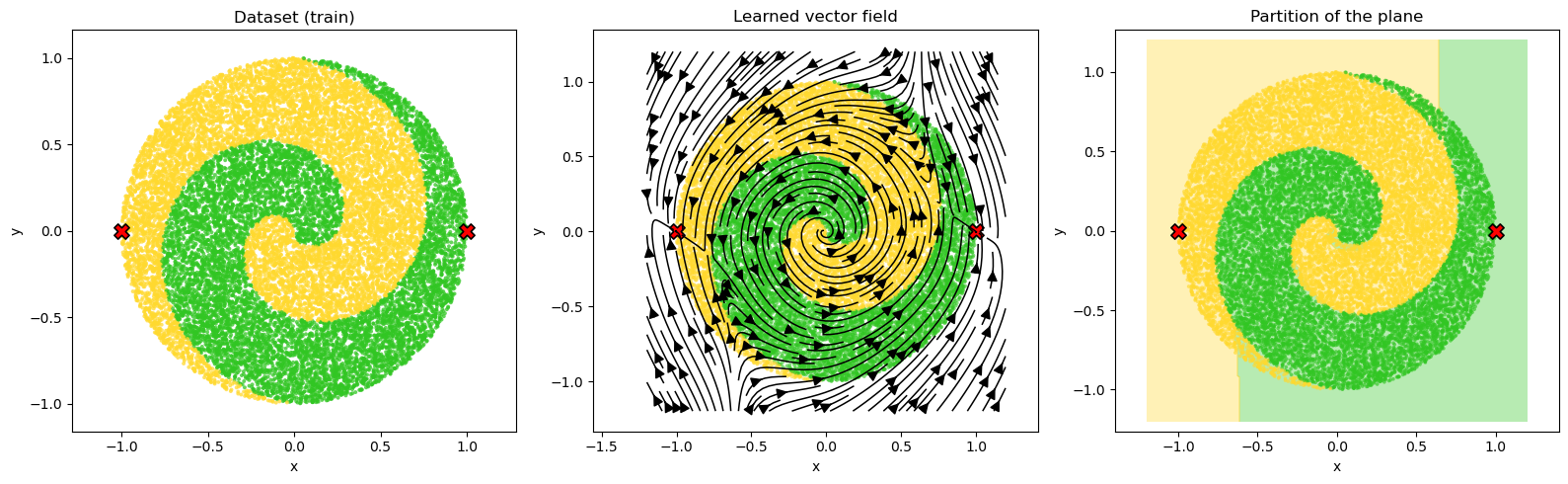}
    \caption{Two-spirals dynamical classification with planted equilibria (ConsNODEs).
    (\emph{Left}) Training set (two interleaving spirals) with planted equilibria marked by red crosses.
    (\emph{Center}) Learned vector field visualized via streamlines, showing how trajectories are routed toward the
    corresponding class equilibrium.
    (\emph{Right}) Induced partition of the plane obtained by integrating each initial condition to time $T$ and applying
    the nearest-equilibrium decoder \eqref{eq:nearest_attractor_rule}.}
    \label{fig:spirals_results}
\end{figure}

Figure~\ref{fig:spirals_results} summarizes the outcome of the analysis.
The learned vector field reshapes the plane so that initial conditions belonging to different spirals are driven,
within the finite horizon $T$, into distinct basins of attraction associated with the two planted equilibria.
In this low-dimensional setting the classifier becomes directly interpretable: the decision boundary corresponds to
the separatrix induced by the flow and can be visualized by propagating a grid of initial conditions to time $T$
and assigning each point to its corresponding class equilibrium. 

\begin{figure}[H]
    \centering
    \includegraphics[width=0.7\linewidth]{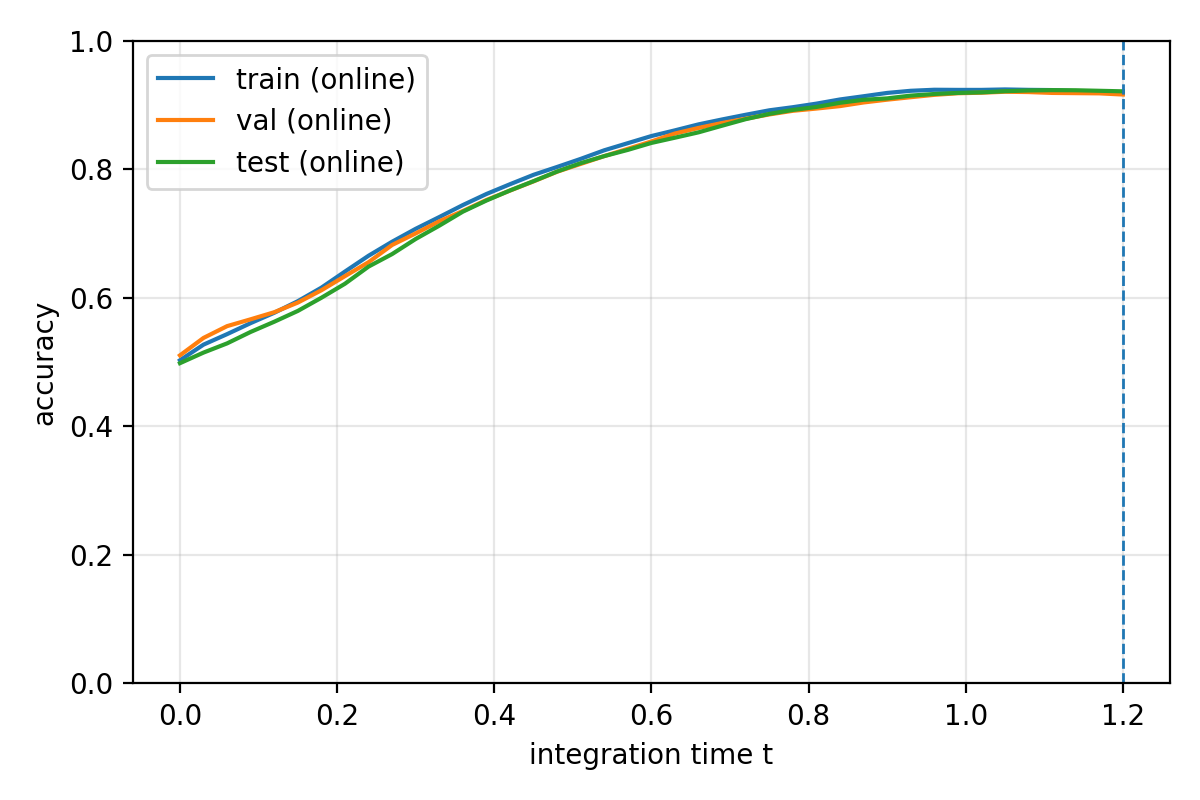}
    \caption{Accuracy of ConsNODEs on the two spirals as a function of the integration time, for the training, validation and test sets. The dashed vertical line marks the training horizon $T=1.2$. Starting from chance level at $t=0$, the accuracy grows essentially monotonically and reaches, at $t\simeq0.9$, a plateau within $0.5$ percentage points of its value at $T$.}
    \label{fig:spirals_acc_time}
\end{figure}

The accuracy of the nearest-equilibrium decoder at time $T$ is $91.9\,\%$ on the training set, $91.7\,\%$ on the validation set and $92.2\,\%$ on the test set. Since the decoder can be applied at any intermediate time $t\le T$, we can follow how the classification builds up along the integration: Fig.~\ref{fig:spirals_acc_time} shows the accuracy as a function of $t$.

\begin{figure}[H]
    \centering
    \includegraphics[width=0.55\linewidth]{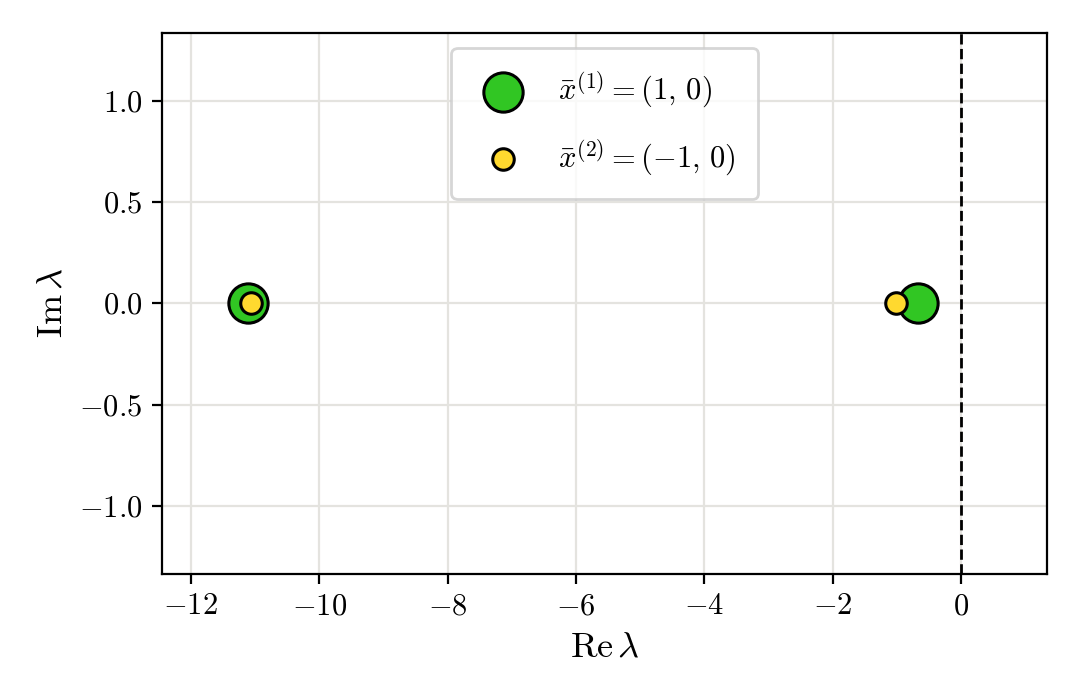}
    \caption{Two spirals: eigenvalues of the Jacobian $J^{(\ell)}$, Eq.~\eqref{eq:jacobian_planted}, at the two planted equilibria of the trained model of Fig.~\ref{fig:spirals_results}, in the complex plane (horizontal axis $\operatorname{Re}\lambda$, vertical axis $\operatorname{Im}\lambda$). Each equilibrium has $n=2$ eigenvalues, drawn with the color of its class in Fig.~\ref{fig:spirals_results} (green, larger markers: $\bar x^{(1)}$; yellow, smaller markers: $\bar x^{(2)}$). The dashed line marks $\operatorname{Re}\lambda=0$. The four eigenvalues are real and negative: both equilibria are stable nodes.}
    \label{fig:spirals_jacobian_spectrum}
\end{figure}

By construction the two planted points are equilibria of the learned field; their stability is decided by the Jacobian $J^{(\ell)}$ of Eq.~\eqref{eq:jacobian_planted}, which is here a $2\times2$ matrix (for the sigmoid, $f'(z)=f(z)\,[1-f(z)]$ in $D_\ell$). Its eigenvalues for the trained model are shown in Fig.~\ref{fig:spirals_jacobian_spectrum}. They are real and negative, $\lambda=-0.67$ and $-11.11$ at $\bar x^{(1)}$, $\lambda=-1.02$ and $-11.06$ at $\bar x^{(2)}$: both planted equilibria are stable nodes, with spectral abscissa $\alpha_1=-0.67$ and $\alpha_2=-1.02$ (Sec.~\ref{sec:jacobian_stability}). They are also stable fixed points of the explicit Euler map $x\mapsto x+\Delta t\,F_\theta(x)$ that is iterated in practice, since $|1+\Delta t\,\lambda|\le0.98$ for all the eigenvalues. In this experiment the equilibria are already stable at the initialization of the parameters ($\lambda=-0.51$ and $-1.16$ at both points), so that the training does not have to create their stability, at variance with the cases discussed in Secs.~\ref{subsec:unstable_to_stable} and \ref{subsec:models_compared}.

\subsection{Oscillation of classification}
\label{subsec:oscillation_classification}

The accuracy-versus-time curves of Figs.~\ref{fig:spirals_acc_time} and~\ref{fig:fmnist_acc_time} average over the whole dataset and therefore do not reveal what happens to a single trajectory. The nearest-equilibrium decoder \eqref{eq:nearest_attractor_rule} can however be applied at any intermediate time $t\in[0,T]$,
\begin{equation}
\widehat y_i(t) := \arg\min_{\ell\in\{1,\dots,C\}} \bigl\|x_i(t)-\bar x^{(\ell)}\bigr\|_2 ,
\label{eq:yhat_time}
\end{equation}
where $x_i(t)$ denotes the trajectory issued from the $i$-th input. It is then natural to ask whether the label assigned to a given input changes monotonically in time, or whether it can switch back and forth before the trajectory settles on its final equilibrium. To address this question we track, along each trajectory, how much closer the state is to the correct equilibrium than to any wrong one.

Let $y_i$ be the true class of input $i$. For every time $t$ we compute the distances to all the planted equilibria, $d_{i,\ell}(t):=\|x_i(t)-\bar x^{(\ell)}\|_2$, and we denote by
\begin{equation}
d^{+}_i(t):=d_{i,y_i}(t),
\qquad
d^{-}_i(t):=\min_{\ell\neq y_i} d_{i,\ell}(t),
\end{equation}
the distance from the correct equilibrium and the distance from the closest wrong one, respectively. The quantity $d^{-}_i-d^{+}_i$ is positive if and only if $\widehat y_i(t)=y_i$. We work with its normalized version (which we term \emph{normalized margin})
\begin{equation}
\tilde m_i(t) := \frac{d^{-}_i(t)-d^{+}_i(t)}{d^{-}_i(t)+d^{+}_i(t)}\;\in\;[-1,1].
\label{eq:mtilde}
\end{equation}
This choice is motivated by the following properties.
\begin{enumerate}
\item[(i)] \emph{It does not alter the decision rule.} $\tilde m_i(t)$ has the same sign as $d^{-}_i(t)-d^{+}_i(t)$: it is positive if and only if the rule \eqref{eq:yhat_time} returns the correct class, and it vanishes on the boundary of the region where the correct class is returned. Every switch of the prediction between the correct class and a wrong one is therefore a zero-crossing of $\tilde m_i(t)$, while the magnitude of $\tilde m_i(t)$ quantifies how far the state is from an ambiguous assignment.
\item[(ii)] \emph{It is bounded.} Since $d^{\pm}_i\ge 0$, one has $|d^{-}_i-d^{+}_i|\le d^{-}_i+d^{+}_i$, hence $-1\le\tilde m_i\le 1$. The value $+1$ is attained when the trajectory sits exactly on the correct equilibrium ($d^{+}_i=0$), the value $-1$ when it sits on a wrong one ($d^{-}_i=0$).
\end{enumerate}

The number of sign changes of $\tilde m_i(t)$ along the (discretized) integration interval counts how many times the prediction for input $i$ switches between the correct class and a wrong one. We say that the classification of input $i$ \emph{oscillates} when $\tilde m_i(t)$ crosses zero at least twice, i.e.\ when such a switch occurs at least twice along the trajectory. A single crossing is not counted, as it simply signals a trajectory that starts on the wrong side of the decision boundary of the nearest-equilibrium rule and eventually reaches the correct basin. Among the oscillating trajectories we further select those that end close to the correct equilibrium, $\tilde m_i(t_{\rm end})\ge 0.95$ (which implies that they are correctly classified, $\tilde m_i(t_{\rm end})>0$), where the flow is integrated up to $t_{\rm end}=6=5T$, well beyond the training horizon $T$. These are the trajectories that, after having changed label back and forth, actually converge to the correct attractor and no longer change label: the final classification is stable.

\begin{figure}[H]
    \centering
    \begin{subfigure}{0.48\linewidth}
        \centering
        \includegraphics[width=\linewidth]{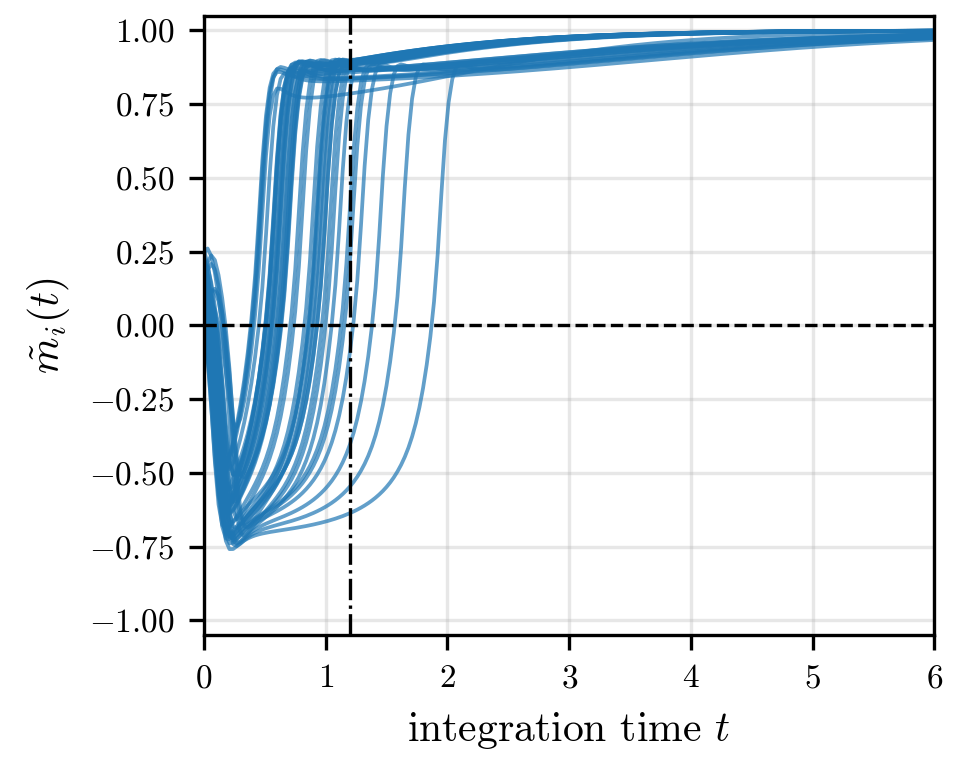}
        \caption{Training set}
        \label{fig:oscillation_plateau_spirals_train}
    \end{subfigure}
    \hfill
    \begin{subfigure}{0.48\linewidth}
        \centering
        \includegraphics[width=\linewidth]{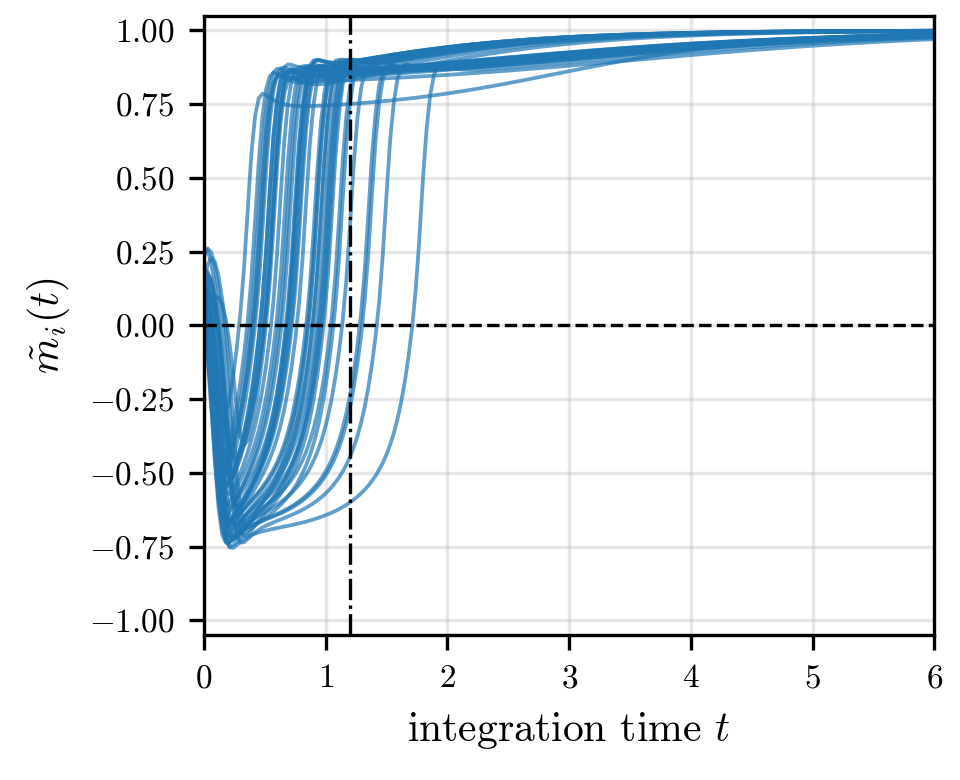}
        \caption{Test set}
        \label{fig:oscillation_plateau_spirals_test}
    \end{subfigure}
    \caption{Oscillation of the classification for the two-spirals dataset, on the training set (a) and on the test set (b).
    The flow is integrated up to $t=6$, well beyond the training horizon $T=1.2$ (dash-dotted vertical line).
    Each curve is the normalized margin $\tilde m_i(t)$, Eq.~\eqref{eq:mtilde}, of a single input as a function of the integration time $t$.
    The dashed line $\tilde m=0$ is the decision boundary of the nearest-equilibrium rule: $\tilde m_i>0$ ($\tilde m_i<0$) means that at time $t$ the input is assigned to the correct (to the wrong) class.
    Only trajectories that oscillate and converge to the correct equilibrium are shown: they cross $\tilde m=0$ at least twice (all the displayed ones exactly twice) and have $\tilde m_i(6)\ge 0.95$.
    They are selected among $1000$ randomly drawn inputs of each set: $86$ (training) and $81$ (test) of them satisfy these requirements, and $40$ are displayed in each panel.
    All the curves start on the correct side of the boundary ($0<\tilde m_i(0)\lesssim 0.3$), are pushed to the wrong side at early times (first crossing at $t\lesssim 0.2$, minimum $\tilde m_i\simeq-0.75$ at $t\simeq 0.15$--$0.35$), and re-enter the correct region at $t\simeq 0.3$--$1.9$; in each panel $4$ of the $40$ curves re-enter it only after $T$, i.e.\ they are still misclassified at the training horizon.
    Afterwards $\tilde m_i$ keeps increasing slowly, from $\tilde m_i(T)\lesssim 0.9$ to $\tilde m_i(6)\ge 0.97$: the trajectories are still relaxing towards the correct equilibrium at $t=T$ and reach it ($\tilde m_i=1$) only for $t\gg T$. This shows that heavily nested data points may require an integration time longer than the training horizon to actually reach the attractor. One could in principle combine both criteria: the nearest-equilibrium rule gives an immediate, possibly still transient, classification, while $\tilde m_i(t)$ measures how close the trajectory is to the attractor, so that a trajectory can be considered fully settled once it reaches a plateau close to $\tilde m_i\simeq 1$. It is nonetheless remarkable that the vast majority of trajectories are already correctly classified well before the training horizon, even though they have not yet reached the attractor.}
    \label{fig:oscillation_plateau_spirals}
\end{figure}

Figure~\ref{fig:oscillation_plateau_spirals} displays these trajectories for the two-spirals experiment of Sec.~\ref{subsec:spirals_toy}, for the training and for the test set.

\subsection{Distribution of the final decision times}
\label{subsec:decision_times}

\begin{figure}[htbp]
    \centering
    \begin{subfigure}{0.48\linewidth}
        \centering
        \includegraphics[width=\linewidth]{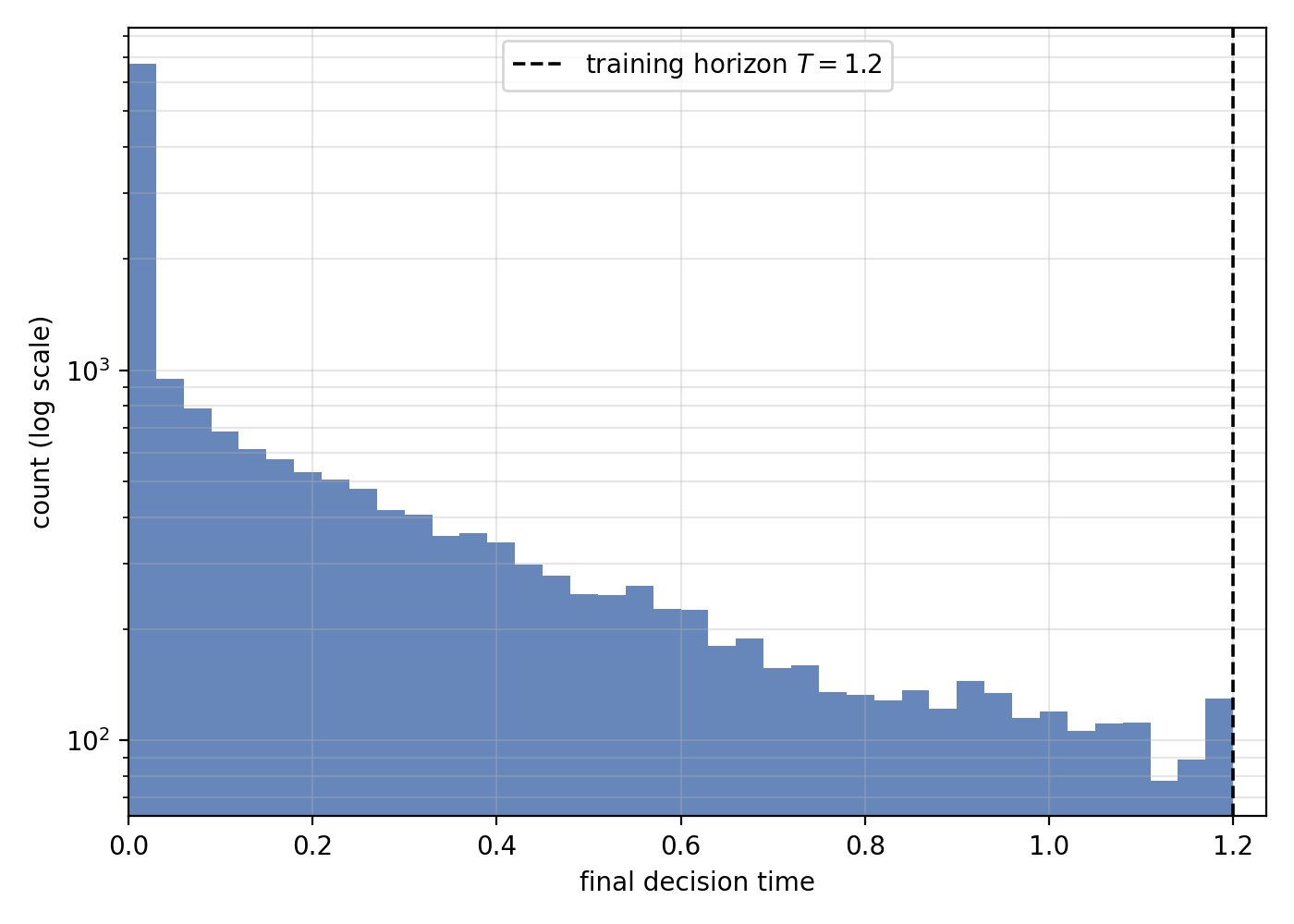}
        \caption{Training set}
        \label{fig:decision_time_spirals_train}
    \end{subfigure}
    \hfill
    \begin{subfigure}{0.48\linewidth}
        \centering
        \includegraphics[width=\linewidth]{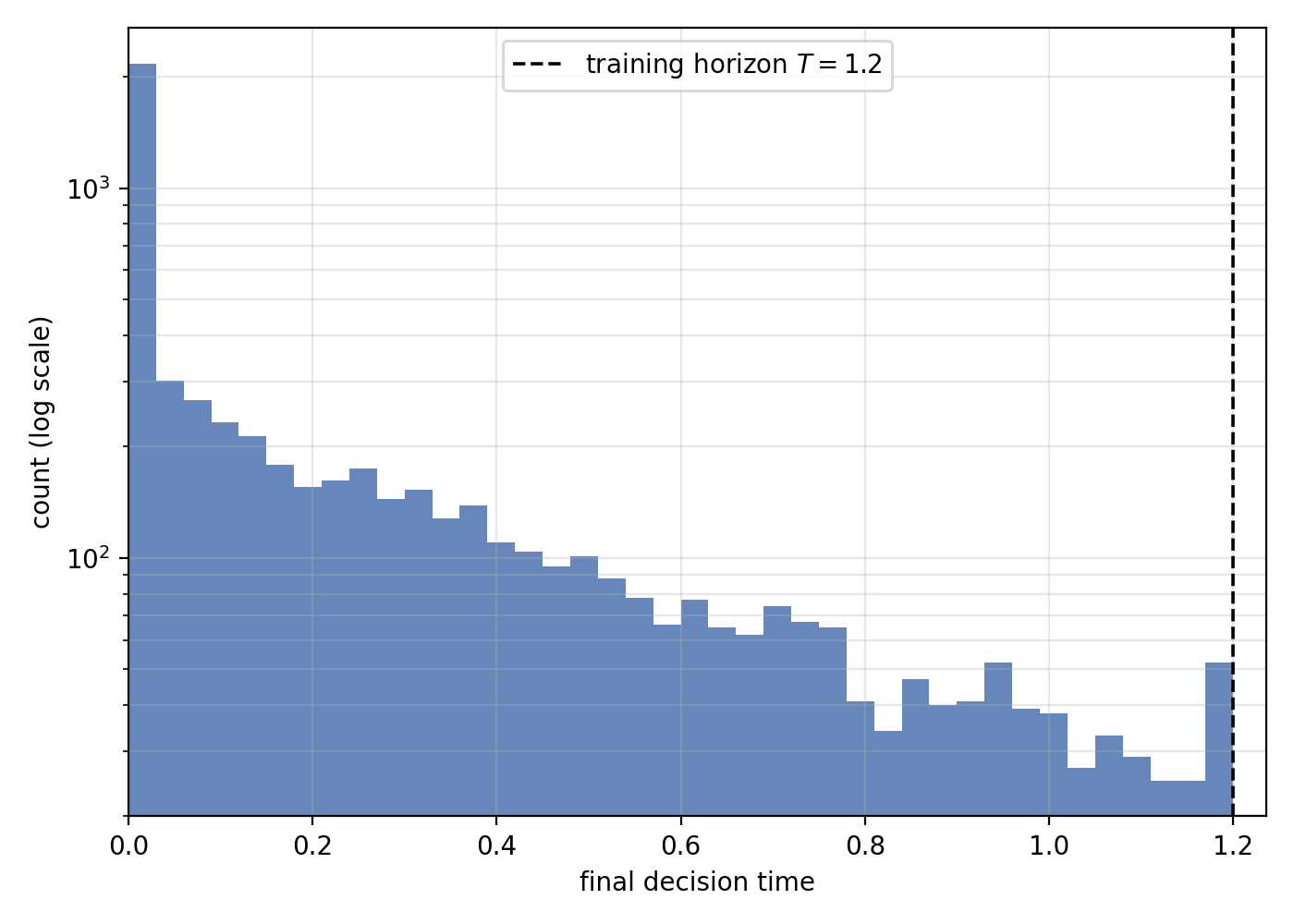}
        \caption{Test set}
        \label{fig:decision_time_spirals_test}
    \end{subfigure}
    \caption{Distribution of the final decision times $\tau_i$ for the two-spirals dataset, on the training set (a, $18{,}000$ points) and on the test set (b, $6{,}000$ points).
    $\tau_i$ is the last time at which the prediction of the nearest-equilibrium rule, Eq.~\eqref{eq:yhat_time}, changes along $[0,T]$ ($\tau_i=0$ if it never changes); it is computed on the discretized trajectory ($\Delta t=0.03$) and the histogram has $40$ bins of width $\Delta t$.
    The vertical axis is logarithmic, and the dashed line marks the training horizon $T=1.2$.}
    \label{fig:decision_time_spirals}
\end{figure}

The previous subsection follows a handful of selected trajectories. To see how the switching of the label is distributed over the whole dataset, we assign to each input a \emph{final decision time}
\begin{equation}
\tau_i := \max\bigl\{t_k\in\{\Delta t,2\Delta t,\dots,T\} \;:\; \widehat y_i(t_k)\neq\widehat y_i(t_{k-1})\bigr\},
\end{equation}
i.e.\ the last time, along the integration interval $[0,T]$ of the training, at which the prediction \eqref{eq:yhat_time} of the nearest-equilibrium rule changes (equivalently, for the two classes considered here, the last sign change of $\tilde m_i(t)$, Eq.~\eqref{eq:mtilde}); we set $\tau_i=0$ if the prediction never changes. With more than two classes the prediction can also switch between two wrong classes, without a sign change of $\tilde m_i(t)$; the definition above counts these switches as well (see Appendix~\ref{sec:fmnist_per_sample}). After $\tau_i$ the label of input $i$ is frozen, whether it is the correct one or not. Figure~\ref{fig:decision_time_spirals} shows the histogram of $\tau_i$ for the two-spirals experiment of Sec.~\ref{subsec:spirals_toy}, on the whole training set ($18{,}000$ points) and on the whole test set ($6{,}000$ points), with a logarithmic vertical axis.

The two histograms are almost indistinguishable: the distribution of the decision times is the same on unseen data as on the data used for training. The first bin contains exactly the inputs whose label never changes ($\tau_i=0$: $37\%$ of the training and $36\%$ of the test inputs), and is about seven times higher than the second one. Overall the decision is taken early: the median of $\tau_i$ is $0.09$ (training) and $0.12$ (test), that is roughly $T/10$. Beyond the first bin the counts decay approximately linearly in the logarithmic scale, i.e.\ roughly exponentially in $\tau$, so that late decisions are rare but not absent: $5.5\%$ (training) and $5.3\%$ (test) of the inputs have $\tau_i>0.9$, i.e.\ they settle only in the last quarter of the interval. The slight upturn in the last bin collects the trajectories whose label is still changing at the training horizon (at the very last step for $0.3\%$ of the training and $0.5\%$ of the test inputs): for these the classification is not yet stable at $T$.

\subsection{Classifying with interacting populations: the two spirals}
\label{subsec:two_pop_spirals}

\begin{figure}[H]
    \centering
    \begin{subfigure}{0.45\linewidth}
        \centering
        \includegraphics[width=\linewidth]{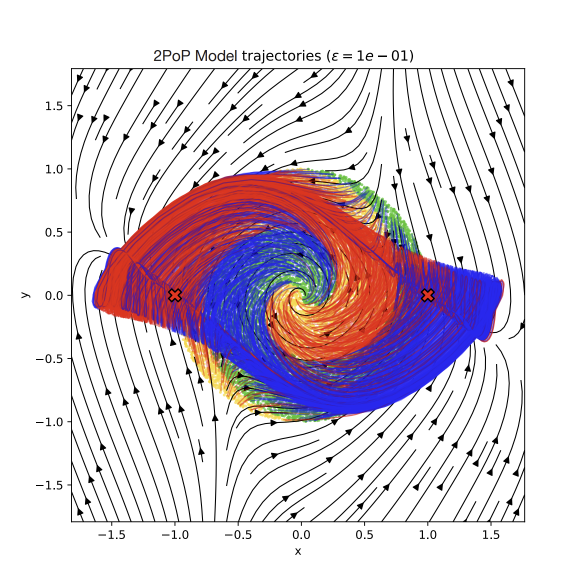}
        \caption{$\varepsilon= 10^{-1}$}
    \end{subfigure}
    \hfill
    \begin{subfigure}{0.45\linewidth}
        \centering
        \includegraphics[width=\linewidth]{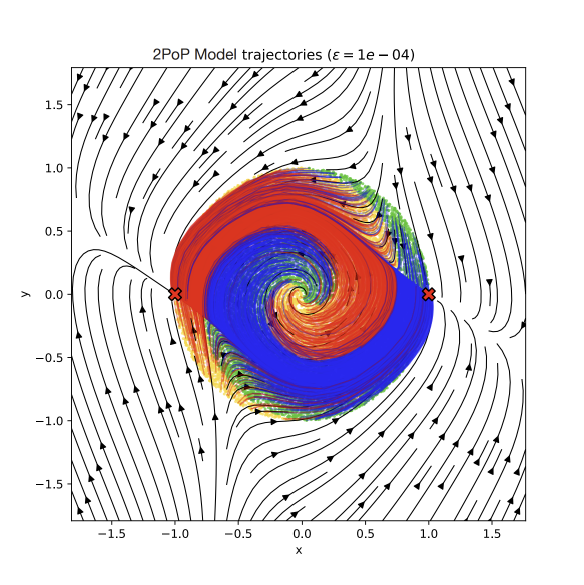}
        \caption{$\varepsilon=10^{-4}$}
    \end{subfigure}

    \caption{Projected trajectories of the coupled model \eqref{eq:coupled_model} in the $x$-space for the two limiting values
    $\varepsilon=10^{-1}$ (a) and $\varepsilon=10^{-4}$ (b) (same trained matrices $(A_1,A_2,b_1,b_2)$ as ConsNODEs).
    The background is that of Fig.~\ref{fig:spirals_results}: the green and the yellow points are the training points of the two classes, the red crosses are the planted equilibria, and the black streamlines are those of the reduced field \eqref{eq:reduced_model_equals_model1}.
    The blue (red) lines are the projected trajectories $x_\varepsilon(t)$, $t\in[0,T]$, issued from $2000$ randomly drawn training points of class~1 (class~2).
    As $\varepsilon$ decreases, the trajectories rapidly align with those induced by the reduced dynamics
    \eqref{eq:reduced_model_equals_model1}. Both systems are integrated with the scheme described in the text, with the step $\Delta t$ of the classifier.}
    \label{fig:eps_trajectories}
\end{figure}

We now come back to the 2PoP model \eqref{eq:coupled_model} introduced in Sec.~\ref{subsec:epsilon_nonzero}. Using the same trained parameters $(A_1,A_2,b_1,b_2)$ obtained for ConsNODEs on the two-spirals dataset
(Sec.~\ref{subsec:spirals_toy}), we integrate the coupled system \eqref{eq:coupled_model} for several values of
$\varepsilon$ and visualize the projected trajectories in the $x$-plane.
The employed numerical integration scheme uses the time step $\Delta t=0.03$ and the horizon $T$ of the classifier. At each step the hidden population is first relaxed towards $A_2x_k+b_2$, with $x_k$ kept fixed,
\[
u_{k+1}=A_2x_k+b_2+e^{-\Delta t/\varepsilon}\big[u_k-(A_2x_k+b_2)\big],
\]
which is the exact solution of the second of Eqs.~\eqref{eq:coupled_model} at fixed $x=x_k$. 
Then $x$ is advanced by one Euler step, $x_{k+1}=x_k+\Delta t\,\big[-x_k+A_1f(u_{k+1})+b_1\big]$; the hidden population starts on the manifold, $u_0=A_2x_0+b_2$. For $\varepsilon\ll\Delta t$ the factor $e^{-\Delta t/\varepsilon}$ vanishes, $u_{k+1}=A_2x_k+b_2$, and the scheme converges to the explicit Euler scheme of the reduced model \eqref{eq:reduced_model_equals_model1} with the step used by the classifier (Sec.~\ref{subsec:spirals_toy}). The experiment below therefore checks that the discretized two-population network reproduces the classifier as $\varepsilon$ decreases. It does not measure the $O(\varepsilon)$ distance between the continuous solutions discussed in Sec.~\ref{subsec:epsilon_nonzero}, which a scheme with a fixed step $\Delta t$ cannot resolve when $\varepsilon\ll\Delta t$.
Figure~\ref{fig:eps_trajectories} shows the two limiting cases $\varepsilon=10^{-1}$ and $\varepsilon=10^{-4}$: as $\varepsilon$ decreases, the projected trajectories $x_\varepsilon(t)$ (drawn in blue and red according to the class of the initial condition) become visually indistinguishable from those generated by the reduced model \eqref{eq:reduced_model_equals_model1},
consistent with the singular-limit argument of Sec.~\ref{subsec:epsilon_nonzero}.

\begin{figure}[H]
    \centering
    \begin{subfigure}{0.49\linewidth}
        \centering
        \includegraphics[width=\linewidth]{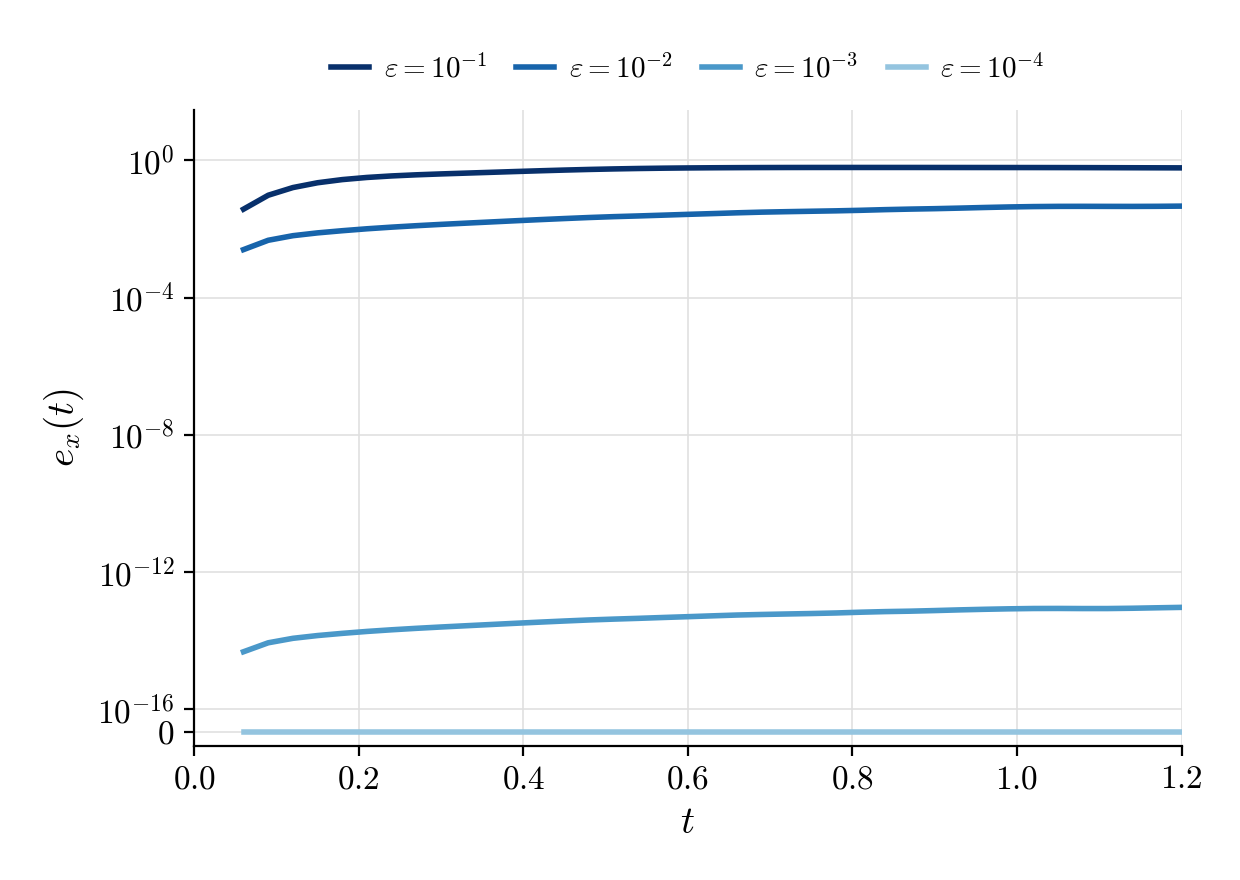}
        \caption{Deviation in $x$-space: $e_x(t)$}
    \end{subfigure}
    \hfill
    \begin{subfigure}{0.49\linewidth}
        \centering
        \includegraphics[width=\linewidth]{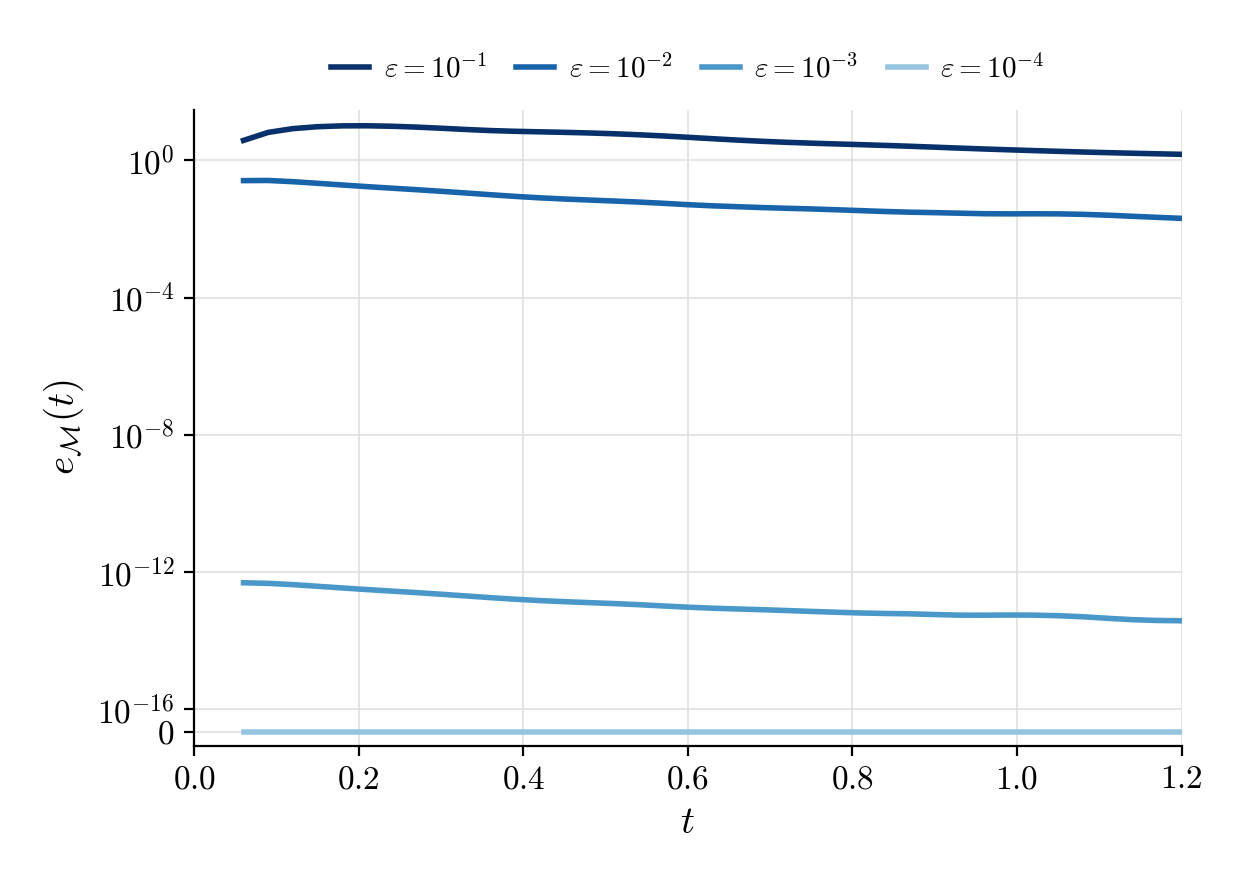}
        \caption{Residual of the relaxation of $u$: $e_{\mathcal M}(t)$}
    \end{subfigure}
    \caption{Diagnostics for the coupled system \eqref{eq:coupled_model} over the horizon $[0,T]$, for $\varepsilon=10^{-1}$, $10^{-2}$, $10^{-3}$ and $10^{-4}$ (averages over $N=2000$ training points, same integration as in Fig.~\ref{fig:eps_trajectories}).
    (\emph{Left}) Mean deviation between the projected coupled trajectory and the trajectory of the Euler scheme of the reduced model \eqref{eq:reduced_model_equals_model1}, with the same step.
    (\emph{Right}) Mean residual of the relaxation of the hidden population towards $A_2x+b_2$, Eq.~\eqref{eq:diagnostics_eps}.
    The vertical axes are logarithmic above $10^{-16}$ and linear below, so that the value zero can be displayed: for $\varepsilon=10^{-4}$ both quantities are zero in double precision. The curves start at $t=2\Delta t$, since both quantities are zero at the first step ($u(0)$ lies on the manifold).
    Both quantities follow the factor $e^{-\Delta t/\varepsilon}$: the plots show that the discretized coupled system reproduces the Euler scheme of the reduced model, not the $O(\varepsilon)$ distance between the continuous solutions (Sec.~\ref{subsec:epsilon_nonzero}).}
    \label{fig:eps_diagnostics}
\end{figure}

To make the convergence more explicit, we monitor two diagnostics along the discrete trajectories, and specifically:
(i) the deviation in $x$ between the coupled and reduced dynamics, and (ii) the residual of the relaxation of the hidden population.
Let $x_k$ denote the iterates of the Euler scheme of the reduced model \eqref{eq:reduced_model_equals_model1}, and $(x^\varepsilon_k,u^\varepsilon_k)$ those of the scheme above for the coupled system, with the same initial condition in $x$.
We define
\begin{equation}
\label{eq:diagnostics_eps}
e_x(t_k) := \mathbb E\big[\|x^\varepsilon_k-x_k\|_2\big],
\qquad
e_{\mathcal M}(t_k) := \mathbb E\big[\|u^\varepsilon_k-(A_2x^\varepsilon_{k-1}+b_2)\|_2\big],
\end{equation}
where $t_k=k\Delta t$ and the expectation is estimated empirically over a set of initial conditions ($N=2000$) drawn from the dataset. $e_{\mathcal M}$ is the distance of the hidden population from the point of $\mathcal M$ towards which it has relaxed during the last step. It is not the distance from $\mathcal M$ of the state $(x^\varepsilon_k,u^\varepsilon_k)$: since $x$ is updated after $u$, the hidden population lags one step behind the visible one, and for $\varepsilon\ll\Delta t$ that distance tends to $\Delta t\,\|A_2\dot x\|$, the discrete counterpart of the displacement $\varepsilon A_2F_\theta(x)$ of the continuous dynamics.
Figure~\ref{fig:eps_diagnostics} reports these quantities over the horizon $[0,T]$ for $\varepsilon=10^{-1}$, $10^{-2}$, $10^{-3}$ and $10^{-4}$, the first and the last being the values of Fig.~\ref{fig:eps_trajectories}. Both decrease rapidly with $\varepsilon$: at $t=T$, $e_x=0.60$ and $e_{\mathcal M}=1.5$ for $\varepsilon=10^{-1}$, $e_x=4.6\times10^{-2}$ and $e_{\mathcal M}=2.0\times10^{-2}$ for $\varepsilon=10^{-2}$, both are of order $10^{-13}$ for $\varepsilon=10^{-3}$, and both are zero in double precision for $\varepsilon=10^{-4}$. These values follow the factor $e^{-\Delta t/\varepsilon}$ of the relaxation step, which is $0.74$, $5.0\times10^{-2}$, $9.4\times10^{-14}$ and $5\times10^{-131}$ for the four values of $\varepsilon$. They measure how closely the discretized coupled system reproduces the Euler scheme of the reduced model, and not the distance between the continuous solutions, which is of order $\varepsilon$ (Sec.~\ref{subsec:epsilon_nonzero}): with the integration scheme of the classifier, the two-population network becomes indistinguishable from the planted Neural-ODE already for $\varepsilon\lesssim10^{-3}$.

Summing up, the model (\ref{eq:coupled_model}) behaves as the modified Neural-ODE with planted fixed points in the limit $\varepsilon\to 0$. Hence, it inherits from the latter the property of universality, as established in \cite{pacificoNEURIPS}. Since the multi-population model also solves the classification task effectively at finite $\varepsilon$, it is tempting to speculate that universality holds in this more general setting as well. This is a
{\bf conjecture} that we leave open for further investigation. In model (\ref{eq:coupled_model}), $A_1$ and $A_2$ are the weighted and signed matrices that couple the two populations $x$ and $u$; the reading that follows from this, and the point at which it enters the stability analysis of the planted equilibria, are discussed in Sec.~\ref{subsec:interpretability}. Finally, when operating under (\ref{eq:coupled_model}) one can in principle overcome the topological constraints that arise when the dimension $n$ of the embedding space is too small for the inherent complexity of the data to be classified. Making the hidden variables explicit, as in (\ref{eq:coupled_model}), provides in fact an additional layer of flexibility: it helps to overcome the limitations intrinsic to an ODE-based classifier, which arise when the trajectories springing from entangled data would have to cross each other (thus violating the Cauchy theorem) in order to reach the designated equilibria. This point will be discussed in a separate contribution devoted to the specific features of model (\ref{eq:coupled_model}).

\subsection{A more interlaced dataset: spirals with two turns}
\label{sec:two_turns}

In Sec.~\ref{subsec:spirals_toy} the two spirals wind once around the origin. Here we make the classes more entangled by letting the two arms wind \emph{twice} around each other: the point $x=r(\cos\theta,\sin\theta)$ of the unit disk is labeled by the sign of $\cos(\theta-4\pi r)$ (instead of $\cos(\theta-2\pi r)$). We draw $N=30{,}000$ balanced points uniformly in the disk and split them in train/validation/test with ratios $0.6/0.2/0.2$, as before, and we plant the same equilibria $\bar x^{(1)}=(1,0)$ and $\bar x^{(2)}=(-1,0)$ at the outer ends of the two arms. The model is the one of Sec.~\ref{subsec:spirals_toy} ($n=2$, sigmoid nonlinearity, explicit Euler scheme with $\Delta t=0.03$, terminal loss \eqref{eq:terminal_regression_loss}, prediction by \eqref{eq:nearest_attractor_rule}), but wider and integrated over a longer interval: $m=600$ hidden units and $80$ steps, i.e.\ $T=2.4$. It is trained for $120$ epochs with Adam (learning rate $1.4\times10^{-3}$); the hyperparameters were selected on the validation accuracy.

\subsubsection{Learned flow and accuracy}
\label{subsec:two_turns_flow}

\begin{figure}[H]
    \centering
    \includegraphics[width=0.98\linewidth]{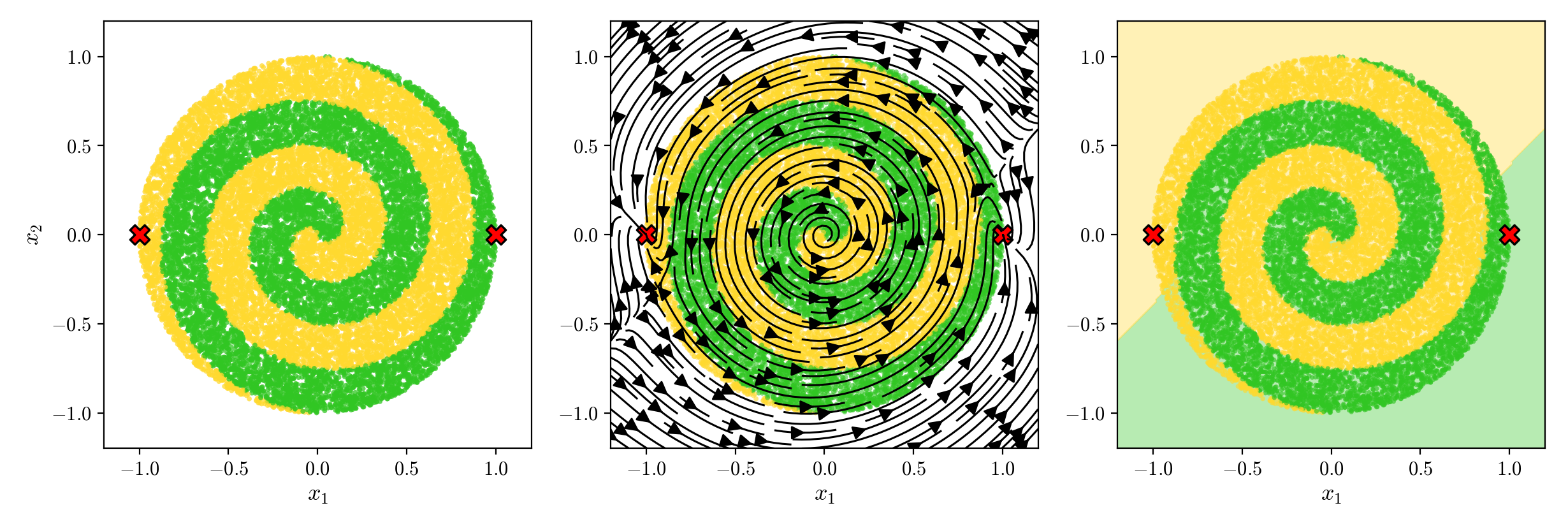}
    \caption{Two spirals with two turns, classified by ConsNODEs with the equilibria planted at $\bar x^{(1)}=(1,0)$ and $\bar x^{(2)}=(-1,0)$ (red crosses).
    (\emph{Left}) Training set.
    (\emph{Center}) Learned vector field visualized via streamlines: the flow winds around the origin following the two arms.
    (\emph{Right}) Induced partition of the plane obtained by integrating each initial condition to time $T=2.4$ and applying the nearest-equilibrium decoder \eqref{eq:nearest_attractor_rule}.
    Test accuracy $93.9\,\%$.}
    \label{fig:spirals_two_turns}
\end{figure}

The accuracy of the nearest-equilibrium decoder at time $T$ is $93.6\,\%$ on the training set, $93.4\,\%$ on the validation set and $93.9\,\%$ on the test set; it is unchanged when the flow is integrated far beyond the training horizon ($93.85\,\%$ on the test set at $t=30$), i.e.\ the classification is not an artifact of stopping at $T$. Figure~\ref{fig:spirals_two_turns} shows the dataset, the learned vector field and the induced partition of the plane. 

\begin{figure}[H]
    \centering
    \includegraphics[width=0.7\linewidth]{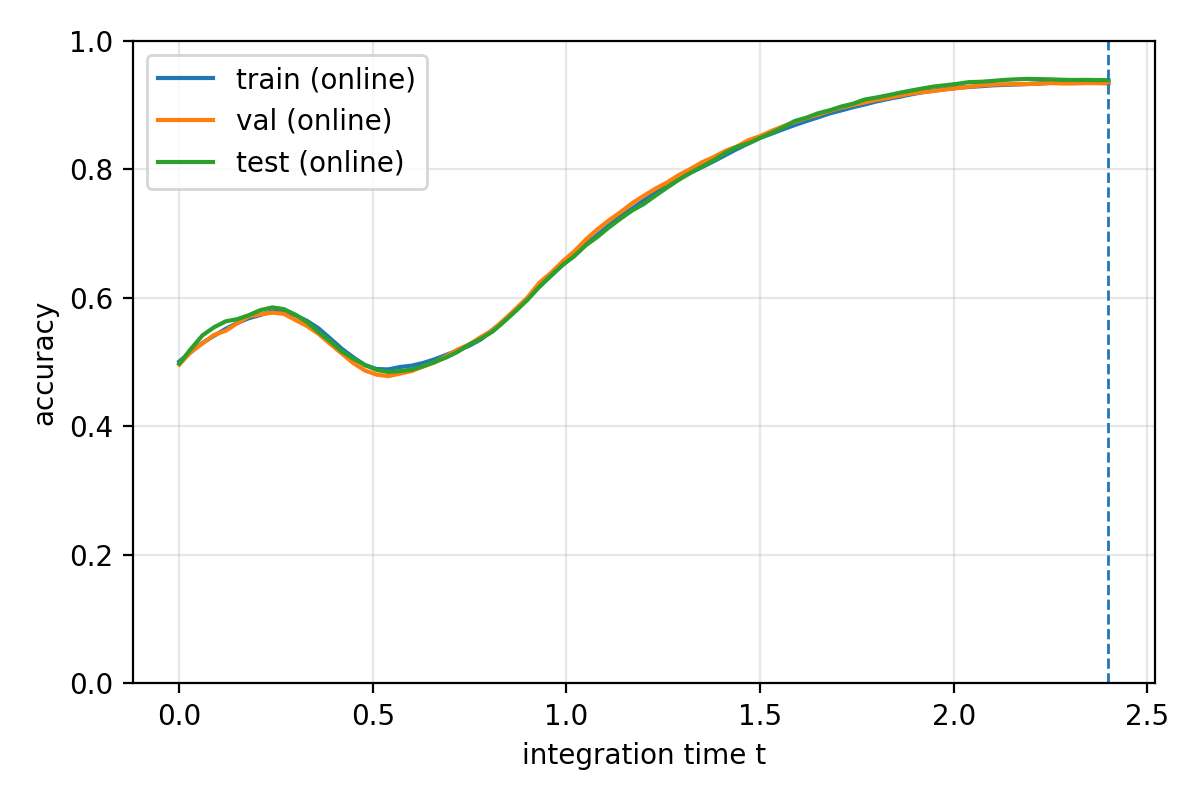}
    \caption{Accuracy of ConsNODEs on the two-turn spirals as a function of the integration time, for the training, validation and test sets. The dashed vertical line marks the training horizon $T=2.4$. Accuracy is non-monotonic at short times, dips below chance level around $t\simeq0.5$, and then rises steadily to the plateau reached at $T$, confirming that the classification is stable rather than an artifact of stopping the integration at $T$.}
    \label{fig:spirals_two_turns_acc_time}
\end{figure}

\begin{figure}[H]
    \centering
    \includegraphics[width=0.55\linewidth]{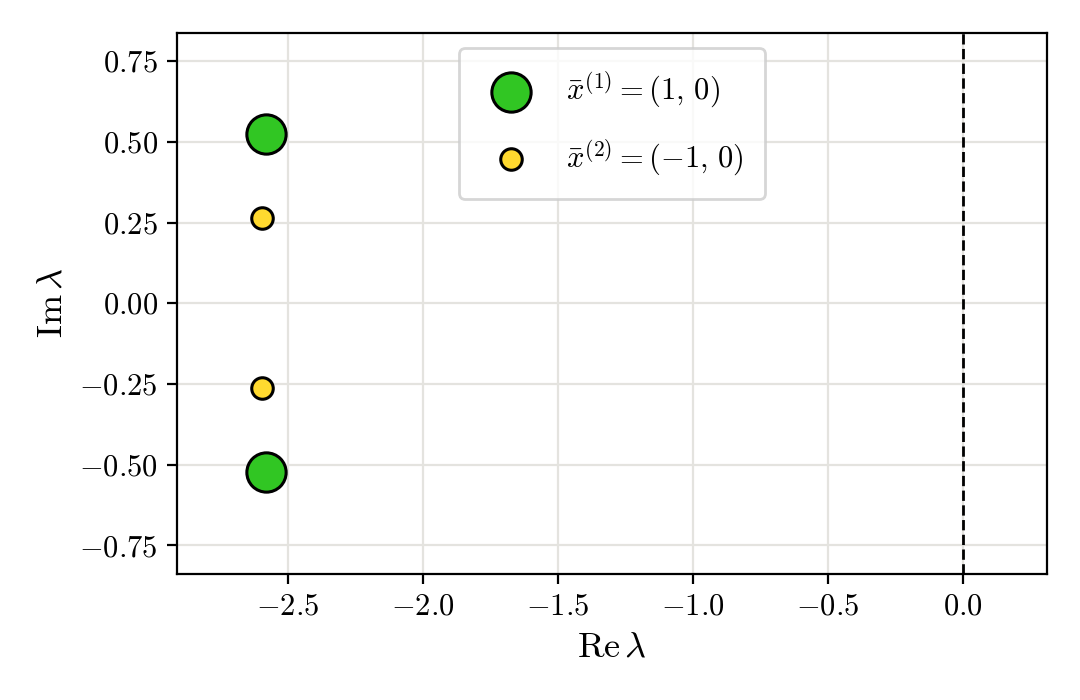}
    \caption{Two spirals with two turns: eigenvalues of the Jacobian $J^{(\ell)}$, Eq.~\eqref{eq:jacobian_planted}, at the two planted equilibria of the trained model of Fig.~\ref{fig:spirals_two_turns}, in the complex plane (horizontal axis $\operatorname{Re}\lambda$, vertical axis $\operatorname{Im}\lambda$). Each equilibrium has $n=2$ eigenvalues, drawn with the color of its class in Fig.~\ref{fig:spirals_two_turns} (green, larger markers: $\bar x^{(1)}$; yellow, smaller markers: $\bar x^{(2)}$). The dashed line marks $\operatorname{Re}\lambda=0$. The eigenvalues are complex conjugate with negative real part: both equilibria are stable foci.}
    \label{fig:spirals_two_turns_jacobian_spectrum}
\end{figure}

The two planted equilibria are attractors of the learned flow also in this case. The eigenvalues of the Jacobian \eqref{eq:jacobian_planted} are complex conjugate with negative real part, $\lambda=-2.58\pm0.52\,i$ at $\bar x^{(1)}$ and $\lambda=-2.60\pm0.26\,i$ at $\bar x^{(2)}$ (Fig.~\ref{fig:spirals_two_turns_jacobian_spectrum}): the equilibria are stable foci, in agreement with the persistence of the classification beyond $T$, and they are stable also for the explicit Euler map ($|1+\Delta t\,\lambda|=0.92$). As for the one-turn spirals, they are already stable at the initialization of the parameters ($\lambda=-0.75$ and $-1.04$ at both points).

\subsubsection{More interlacing, more oscillation}
\label{subsec:two_turns_oscillation}

We repeat the analysis of Sec.~\ref{subsec:oscillation_classification} on this model, counting for each input the zero-crossings of the normalized margin $\tilde m_i(t)$, Eq.~\eqref{eq:mtilde}, along $[0,T]$. The fraction of inputs whose classification oscillates (at least two crossings) grows from $8.4\,\%$ for the one-turn spirals of Sec.~\ref{subsec:spirals_toy} ($T=1.2$) to $39.9\,\%$ for the two-turn ones ($T=2.4$), on the test set; on the training set the values are $8.3\,\%$ and $39.8\,\%$. The switching of the label is therefore much more frequent when the data are more entangled.

\begin{figure}[H]
    \centering
    \begin{subfigure}{0.48\linewidth}
        \centering
        \includegraphics[width=\linewidth]{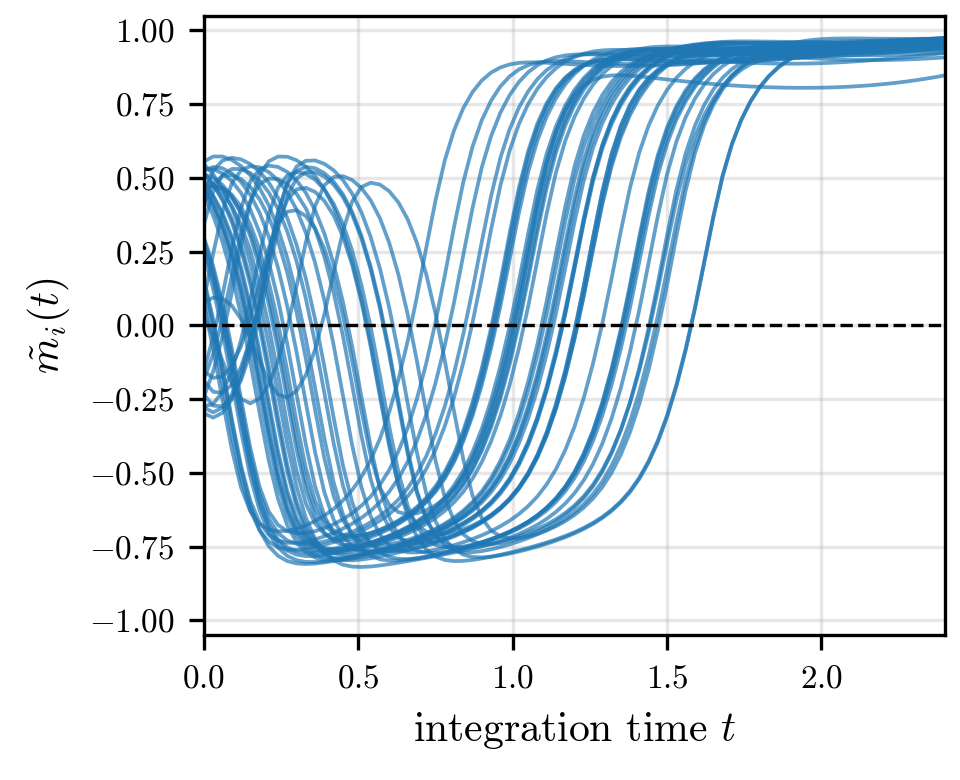}
        \caption{Training set}
        \label{fig:oscillation_plateau_two_turns_train}
    \end{subfigure}
    \hfill
    \begin{subfigure}{0.48\linewidth}
        \centering
        \includegraphics[width=\linewidth]{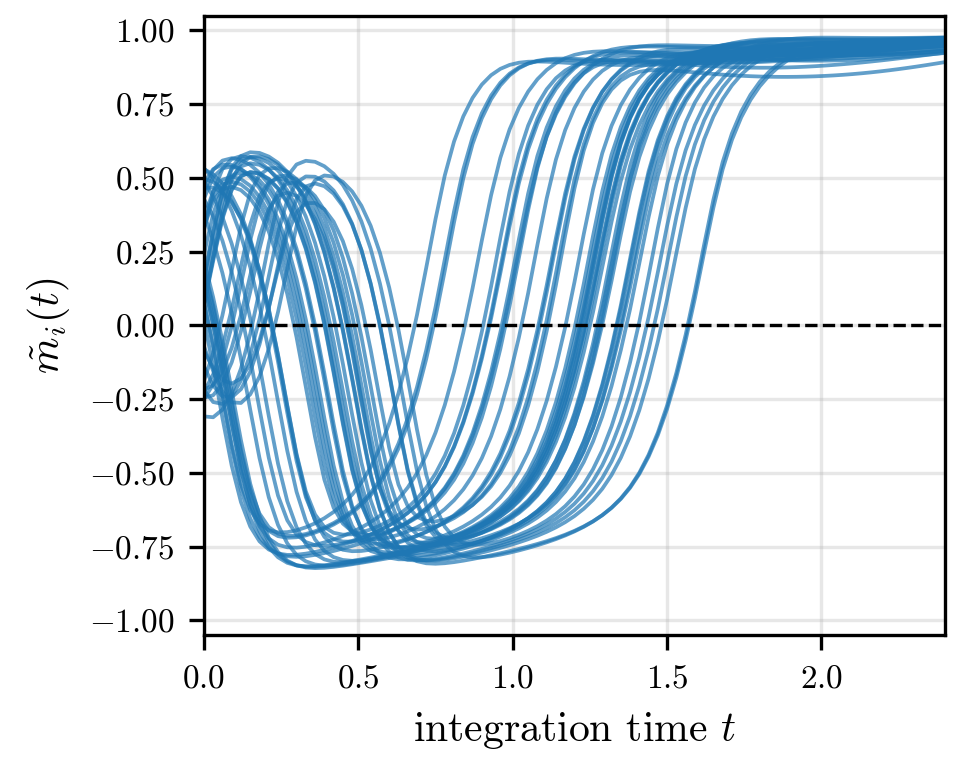}
        \caption{Test set}
        \label{fig:oscillation_plateau_two_turns_test}
    \end{subfigure}
    \caption{Oscillation of the classification for the two-turn spirals, on the training set (a) and on the test set (b).
    Each curve is the normalized margin $\tilde m_i(t)$, Eq.~\eqref{eq:mtilde}, of a single input as a function of the integration time $t\in[0,T]$, $T=2.4$.
    The dashed line $\tilde m=0$ is the decision boundary of the nearest-equilibrium rule.
    Only trajectories that cross $\tilde m=0$ at least twice, are correctly classified at $t=T$ and have settled (the range of $\tilde m_i$ over the last $20\,\%$ of the interval is at most $0.05$) are shown.
    They are selected among $1000$ randomly drawn inputs of each set: $237$ (training) and $237$ (test) satisfy these requirements, and $40$ are displayed in each panel.}
    \label{fig:oscillation_plateau_two_turns}
\end{figure}

Figure~\ref{fig:oscillation_plateau_two_turns} shows the oscillating trajectories that are correctly classified at the end and have settled, for the training and for the test set.

\begin{figure}[H]
    \centering
    \begin{subfigure}{0.48\linewidth}
        \centering
        \includegraphics[width=\linewidth]{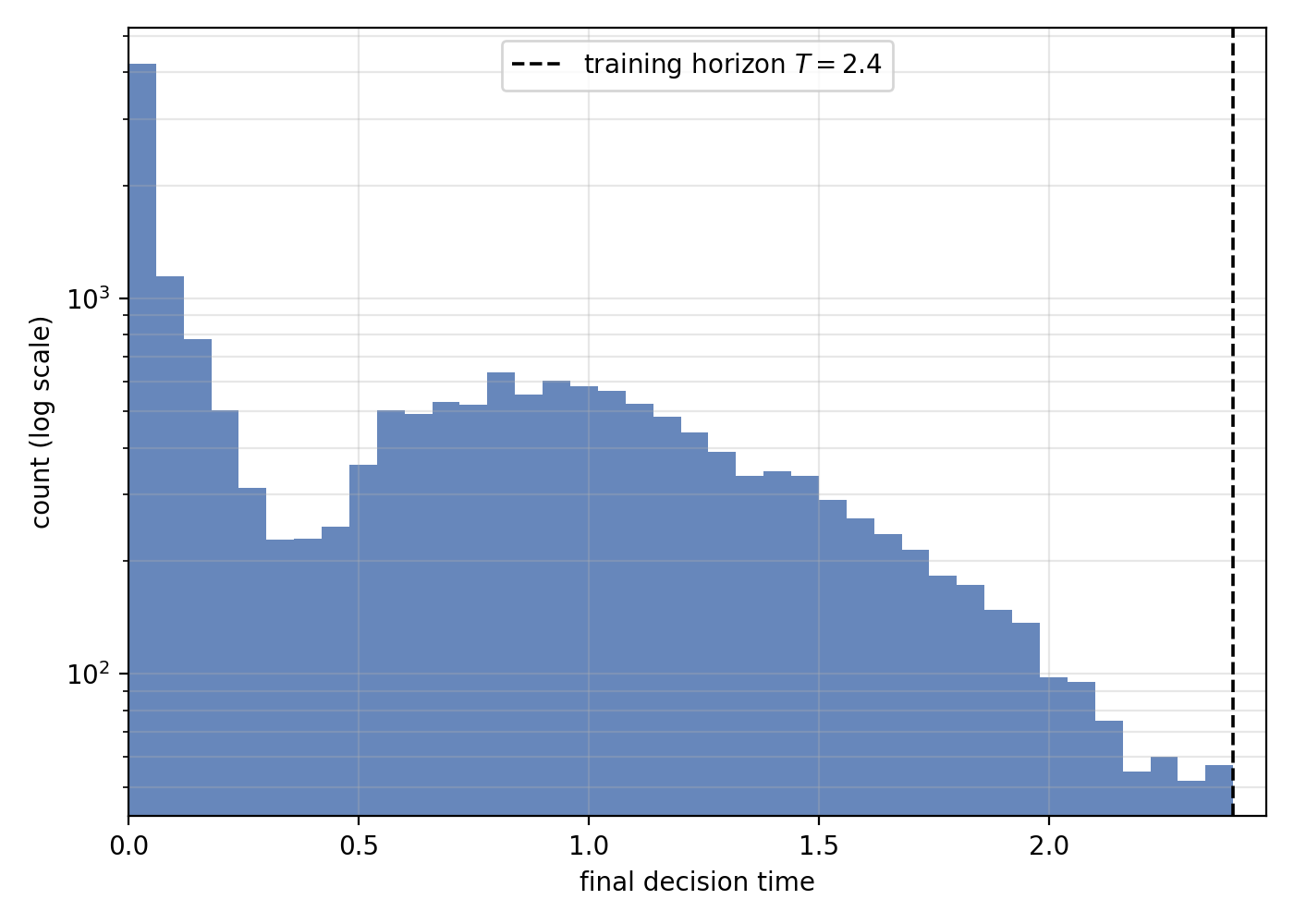}
        \caption{Training set}
        \label{fig:decision_time_two_turns_train}
    \end{subfigure}
    \hfill
    \begin{subfigure}{0.48\linewidth}
        \centering
        \includegraphics[width=\linewidth]{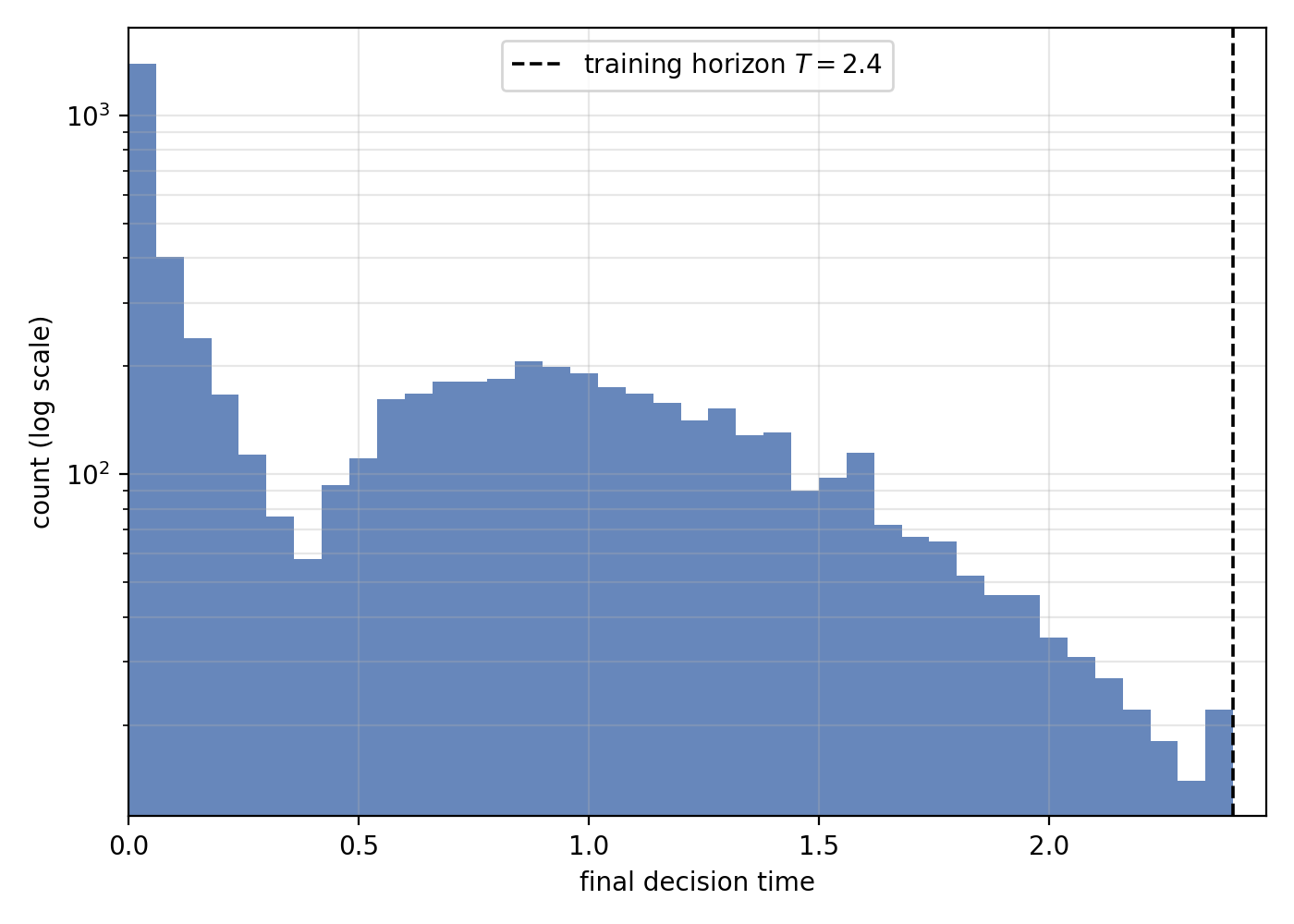}
        \caption{Test set}
        \label{fig:decision_time_two_turns_test}
    \end{subfigure}
    \caption{Distribution of the final decision times $\tau_i$ for the two-turn spirals, on the training set (a, $18{,}000$ points) and on the test set (b, $6{,}000$ points), same definition and binning as Fig.~\ref{fig:decision_time_spirals}.
    The vertical axis is logarithmic, and the dashed line marks the training horizon $T=2.4$.
    Compared with the one-turn spirals of Fig.~\ref{fig:decision_time_spirals}, the distribution is markedly wider and flatter: the first bin ($\tau_i=0$) is still the largest, but the bulk of the decisions is spread over the whole interval up to $T$ rather than concentrated at small $\tau_i$, consistent with the larger fraction of oscillating trajectories reported in Sec.~\ref{subsec:two_turns_oscillation}.}
    \label{fig:decision_time_two_turns}
\end{figure}

As for the one-turn spirals of Sec.~\ref{subsec:decision_times}, Fig.~\ref{fig:decision_time_two_turns} shows the distribution of the final decision times $\tau_i$ for the two-turn spirals, on the whole training and test sets.

\subsection{The topological trap: nested classes and the dimension of the state space}
\label{sec:topological_trap}

The classification rule of Sec.~\ref{subsec:dynamical_classification_principle} asks the flow map $\Psi_T$ of the autonomous field \eqref{eq:generic_node} to drive every initial condition $x_0$ towards the planted equilibrium of its class. The map $\Psi_T$ is however not arbitrary: being the flow of an ordinary differential equation, it inherits its topology from the Cauchy theorem. In this section we make explicit which classification problems this rules out in a state space of a given dimension $n$, and we show on the simplest example, a disk surrounded by an annulus, that the obstruction disappears as soon as one dimension is added. The same limitation of Neural-ODEs, and the same remedy, were pointed out in \cite{dupont2019augmented} for unconstrained Neural-ODEs; here we discuss what it implies for planted-attractor classification, where the target of each class is a prescribed equilibrium.

\subsubsection{A disk surrounded by an annulus in \texorpdfstring{$\mathbb R^2$}{R\texttwosuperior}}
\label{subsec:trap_disk_2d}

We take as class~1 the disk $\{|x|<r_{\rm in}\}$ and as class~2 the annulus $\{r_{\rm in}\le|x|\le r_{\rm out}\}$ of $\mathbb R^2$, with $r_{\rm out}=1$ and $r_{\rm in}=1/\sqrt2$, so that the two regions have the same area. The two classes share the circle $|x|=r_{\rm in}$, and every circle of the annulus encloses the disk. As in Sec.~\ref{subsec:spirals_toy} we sample $N=30{,}000$ balanced points uniformly, split them into train/validation/test with ratios $0.6/0.2/0.2$. The two equilibria $\bar x^{(1)}=(1.5,0)$ and $\bar x^{(2)}=(-1.5,0)$ are planted outside the annulus. The model is that of Sec.~\ref{subsec:spirals_toy} ($n=2$, $m=300$, sigmoid, explicit Euler with $\Delta t=0.03$ and $40$ steps, $T=1.2$), trained on the loss \eqref{eq:terminal_regression_loss} for $40$ epochs with Adam (learning rate $3\times10^{-3}$ with cosine decay, batch size $256$); prediction uses \eqref{eq:nearest_attractor_rule}.

\begin{figure}[H]
    \centering
    \includegraphics[width=0.98\linewidth]{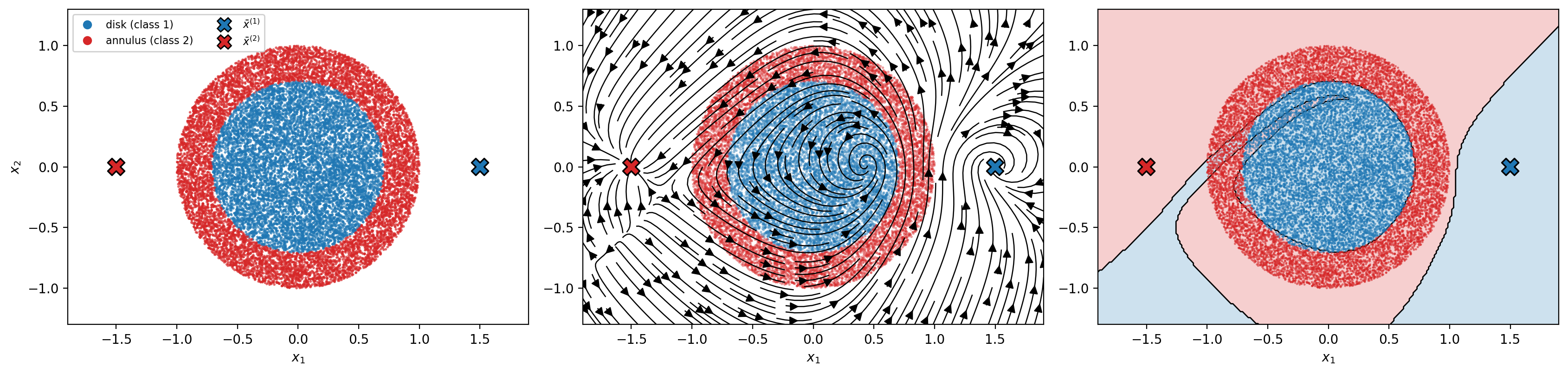}
    \caption{Disk and annulus in $\mathbb R^2$ (ConsNODEs, $n=2$).
    (\emph{Left}) Training set, with the disk (class~1, blue) and the annulus (class~2, red), which share the circle $|x|=r_{\rm in}$, and the planted equilibria $\bar x^{(1)}=(1.5,0)$ and $\bar x^{(2)}=(-1.5,0)$ marked by crosses.
    (\emph{Center}) Learned vector field visualized via streamlines. The disk is captured by an unplanted stable focus located inside the disk, the annulus flows around it towards $\bar x^{(2)}$, and the basin of $\bar x^{(1)}$ lies outside the data.
    (\emph{Right}) Induced partition of the plane obtained by integrating each initial condition to time $T=1.2$ and applying the nearest-equilibrium decoder \eqref{eq:nearest_attractor_rule}.}
    \label{fig:trap_2d}
\end{figure}

The learned field does not send the disk to $\bar x^{(1)}$ (Fig.~\ref{fig:trap_2d}). The disk is captured by an unplanted stable focus inside the disk, at $x\simeq(0.43,0.02)$, whose Jacobian has eigenvalues $-2.0\pm6.8\,i$, while the annulus flows around the disk and converges to $\bar x^{(2)}$ (center panel). The basin of $\bar x^{(1)}$ exists, but it lies outside the data and none of them enters it: in the test set no disk point ends within a distance $0.3$ of $\bar x^{(1)}$ (the median distance is $1.03$), and the same is found in each of the six independent initializations that we trained. The terminal loss \eqref{eq:terminal_regression_loss} on the test set is $0.85\pm0.04$ (mean $\pm$ standard deviation over the six runs), to be compared with $0.069\pm0.005$ for the same model in $\mathbb R^3$, see below. The planar model still reaches a test accuracy of $93.4\pm0.7\,\%$, because the rule \eqref{eq:nearest_attractor_rule} only asks on which side of the bisector of the two equilibria $x(T)$ falls, and the unplanted focus lies on the side of $\bar x^{(1)}$. The residual errors sit around the shared circle, as anticipated: within a distance $0.1$ from it, on either side, $12$--$13\,\%$ of the test points are misclassified, against $2$--$4\,\%$ elsewhere. When the equilibria are moved to $\bar x^{(1)}=(3.5,0)$ and $\bar x^{(2)}=(-1.5,0)$, with $T=3$, so that the whole dataset lies on the side of $\bar x^{(2)}$ of the bisector and this way out is closed, the accuracy itself drops to $77.6\pm8.1\,\%$ over six runs, while the model in $\mathbb R^3$ reaches $96.4\pm5.0\,\%$. The failure is not due to the optimization: trained in the same way on the labels $x_1\ge0$ and $x_1<0$, which carry no nesting, the same planar model reaches both equilibria (test accuracy $99.98\,\%$, and $99.8\,\%$ of the test points end within $0.3$ of the equilibrium of their class).

This example is a concrete instance of the dynamical interpretability introduced in Sec.~\ref{subsec:interpretability}. Judged by its accuracy alone, the planar model would seem to work as designed, with each class sent to its own planted equilibrium. The analysis of the learned field, through its phase portrait and its fixed points, tells a different story. The disk is not carried to $\bar x^{(1)}$: it is trapped by an attractor that was not planted, but created by the training. It is classified correctly only because this attractor happens to lie on the side of $\bar x^{(1)}$ of the bisector, as confirmed by the drop of the accuracy when the equilibria are displaced. Attractors of this kind are the analogue of the spurious states of associative memories \cite{hopfield1982neural, amit1985spin}. These are stable states of the dynamics that coexist with the stored patterns without corresponding to any of them. In associative memories the spurious states are a byproduct of the way the patterns are stored. In the present setting, instead, they encode a property of the data that the accuracy does not reveal. The spurious attractor is, in fact, produced by the topology of the data and not by the planting. Hence the dynamical analysis can bring to light topological obstructions or other features of the data that constrain the classification. This remains possible in high dimension, where neither the phase portrait nor the data can be inspected directly. The test data are integrated beyond the training horizon, and one checks whether they all reach the planted equilibria or other attractors emerge. On Fashion-MNIST no other attractor appears: every test trajectory ends at a planted equilibrium (Appendix~\ref{sec:fmnist_per_sample}).

\subsubsection{Adding a dimension}
\label{subsec:trap_3d}

We now append a third coordinate, equal to zero, to every data point, $x_0\mapsto(x_0,0)\in\mathbb R^3$, and to the equilibria, $\bar x^{(1)}=(1.5,0,0)$ and $\bar x^{(2)}=(-1.5,0,0)$. The data are still a planar disk and a planar annulus, and everything else (the points and their splits, the architecture with $n=3$ and $m=300$, the integration scheme, the optimizer and the number of epochs) is left unchanged. The disk points converge to $\bar x^{(1)}$ ($94\,\%$ of them end within $0.3$ of it, with a median distance of $0.15$), the annulus points converge to $\bar x^{(2)}$ ($89\,\%$), the terminal loss is $0.069\pm0.005$ and the test accuracy is $99.5\pm0.1\,\%$ (six runs).

\begin{figure}[htbp]
    \centering
    \includegraphics[width=0.47\linewidth]{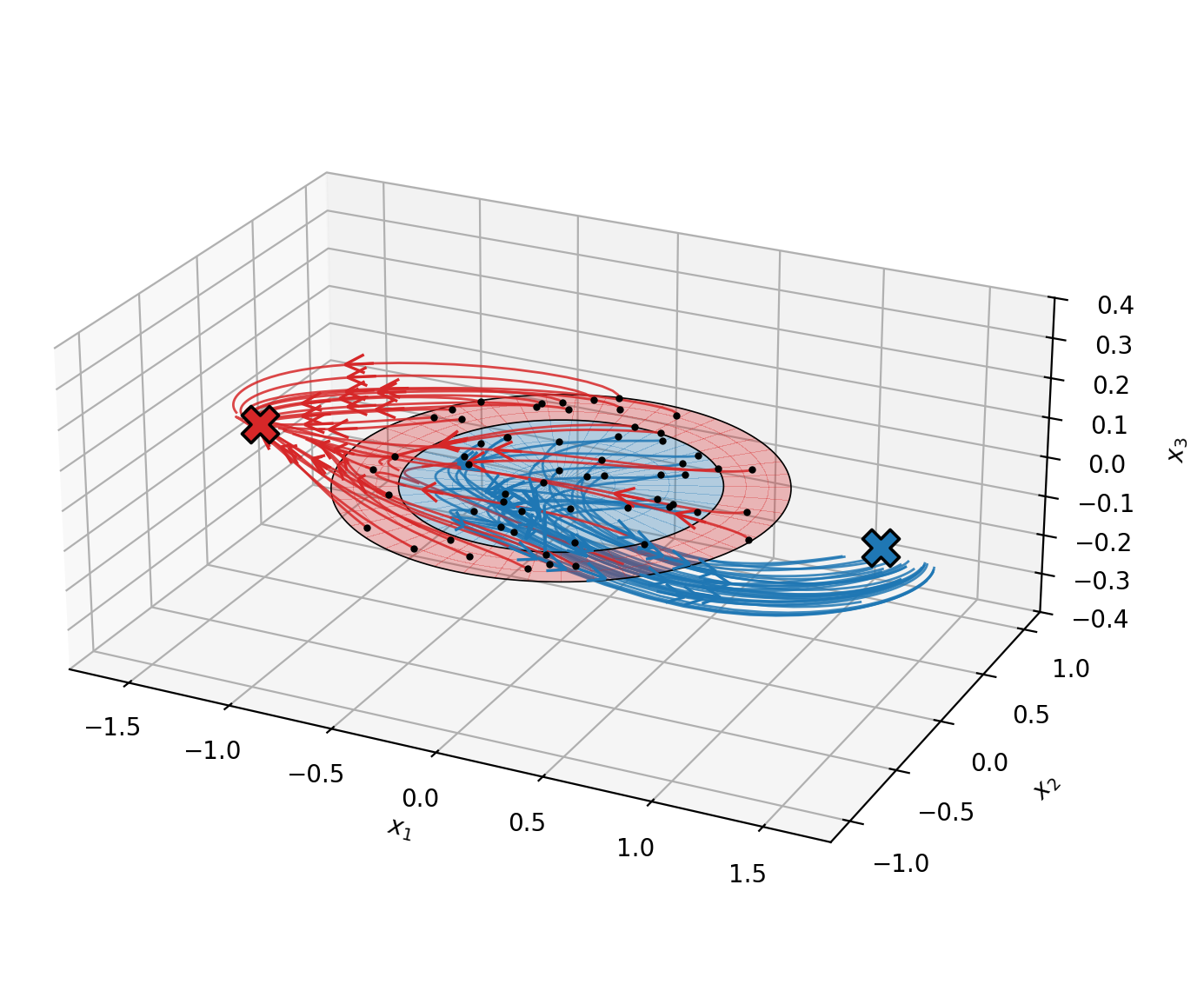}
    \hfill
    \includegraphics[width=0.47\linewidth]{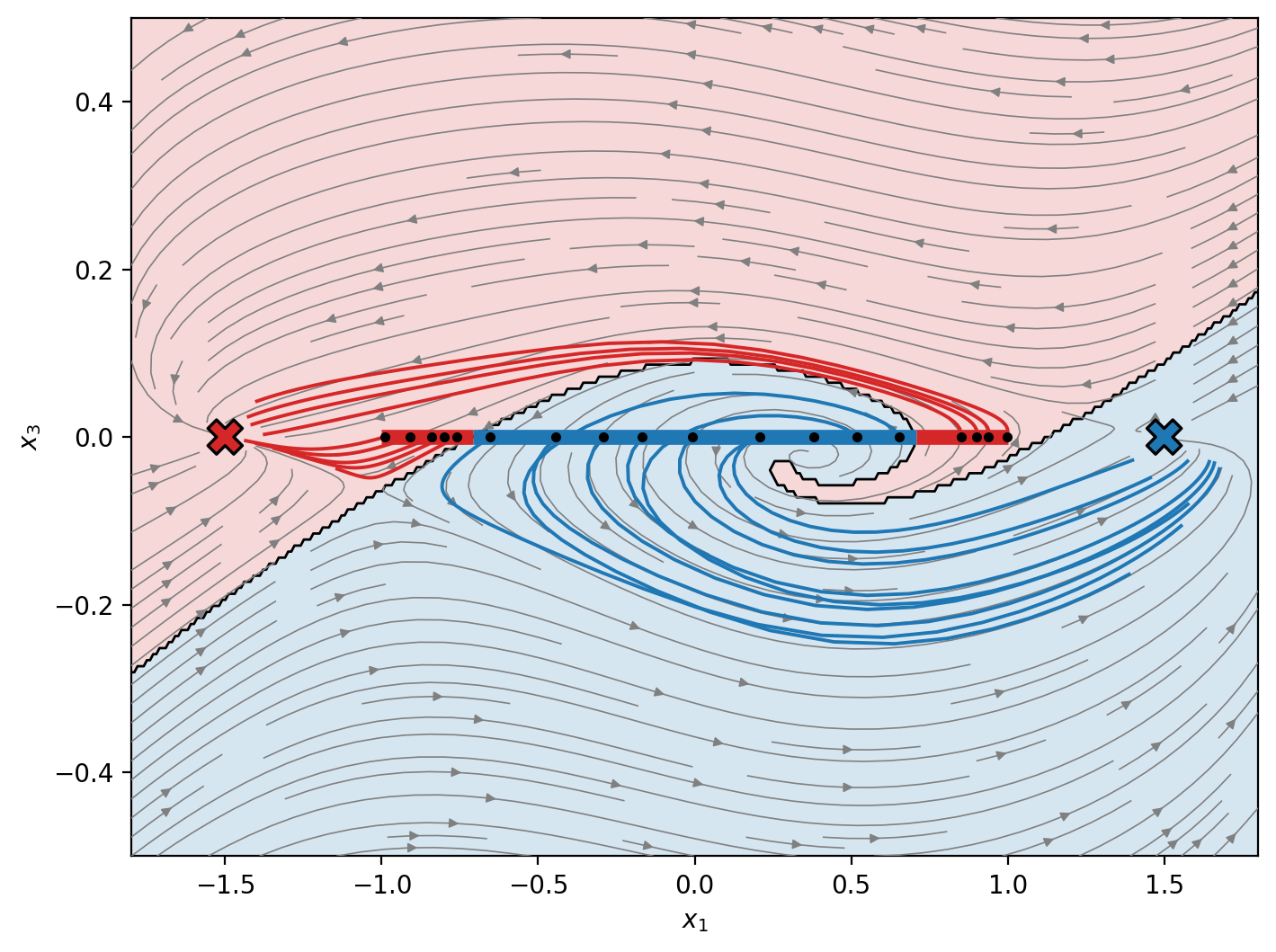}
    \caption{The same problem in $\mathbb R^3$: the data are the planar disk (blue) and annulus (red) in the plane $x_3=0$, and the planted equilibria (crosses) lie in the same plane, at $(\pm1.5,0,0)$.
    (\emph{Left}) Trajectories of $30$ test points per class. The disk and the annulus are drawn as translucent surfaces in the plane $x_3=0$, the arrows give the direction of motion and the $x_3$ axis is stretched by a factor $1.9$. The disk points leave the plane of the data, pass beneath the annulus and converge to $\bar x^{(1)}$; the annulus points move above the plane and converge to $\bar x^{(2)}$.
    (\emph{Right}) Section of the flow with the plane $x_2=0$: streamlines of $(F_1,F_3)$ (gray), basins of attraction of the two equilibria (shading, color of the class assigned by the decoder to the state reached at time $T$), the data cut by the plane (thick segments), and $(x_1,x_3)$ projection of trajectories that start from data points with $|x_2|<0.03$ (a layer of width $0.05$ around the shared circle is left out).}
    \label{fig:trap_3d}
\end{figure}

The mechanism is displayed in Fig.~\ref{fig:trap_3d}. The learned field acquires a component along the new coordinate, of opposite sign on the two classes (its mean over the data is $-0.08$ on the disk and $+0.26$ on the annulus). The disk points leave the plane of the data, pass beneath the annulus and rise onto $\bar x^{(1)}$, while the annulus points move above the plane and go to $\bar x^{(2)}$: the two streams exchange their positions, which in the plane would require them to cross. All the disk trajectories dip below the plane ($x_3<-0.05$), whereas $57\,\%$ of the annulus ones rise above it ($x_3>0.05$). The excursions out of the plane are modest ($|x_3|<0.3$ for all the $6{,}000$ test trajectories), and they are enough. Since the data are invariant under $x_3\to-x_3$, which of the two classes goes below the plane is not imposed by the data: it is decided by the initialization. In the section of Fig.~\ref{fig:trap_3d} (right) the disk lies in the basin of $\bar x^{(1)}$ and the annulus in that of $\bar x^{(2)}$; the basin of $\bar x^{(1)}$ extends below the plane of the data and that of $\bar x^{(2)}$ above it.

\begin{figure}[H]
    \centering
    \includegraphics[width=0.85\linewidth]{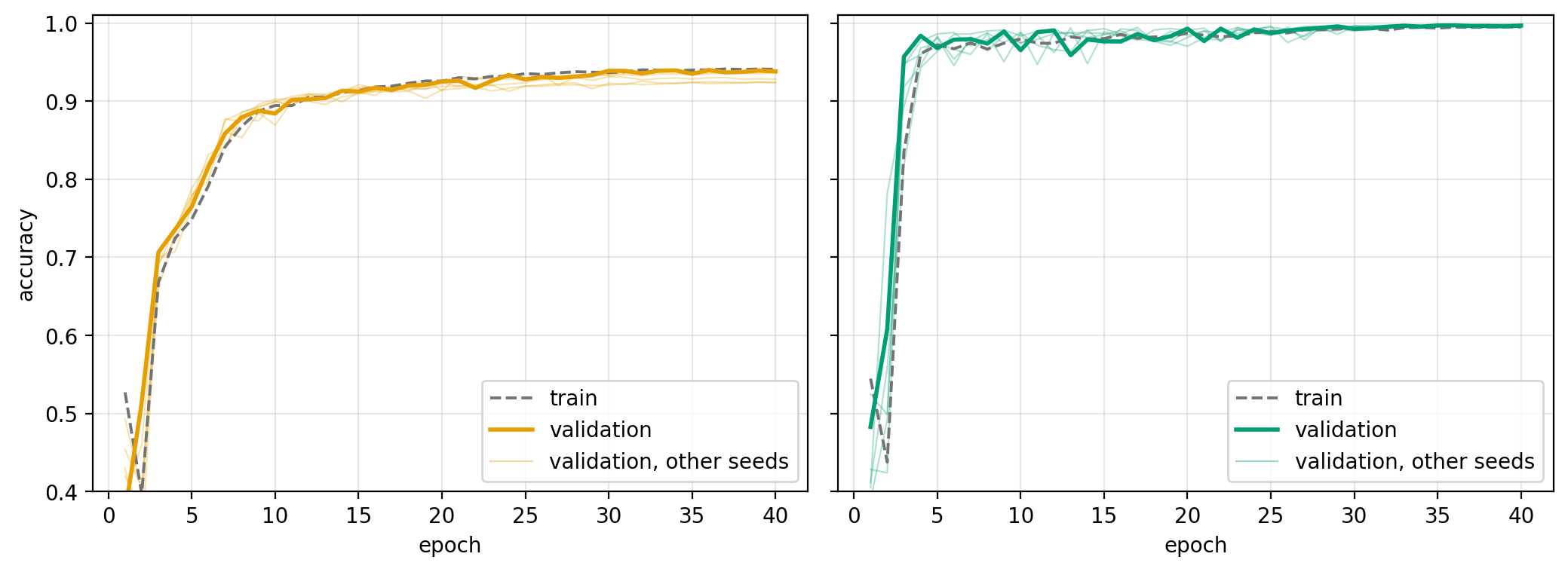}
    \caption{Accuracy as a function of the training epoch for the disk and annulus problem in $\mathbb R^2$ (\emph{left}) and in $\mathbb R^3$ (\emph{right}), with the same points, architecture ($m=300$), integration scheme and optimizer. Dashed: training accuracy; solid: validation accuracy of the run shown in Figs.~\ref{fig:trap_2d} and \ref{fig:trap_3d}; thin lines: validation accuracy of five further runs with different initializations of the parameters. At the last epoch the test accuracy of the run shown is $94.3\,\%$ in $\mathbb R^2$ and $99.6\,\%$ in $\mathbb R^3$.}
    \label{fig:trap_acc}
\end{figure}

Finally, Fig.~\ref{fig:trap_acc} compares the accuracy during training in the two cases. In $\mathbb R^2$ the accuracy saturates at about $94\,\%$, and what limits it is not the optimization but the topology (see the control above); in $\mathbb R^3$, with the same data, architecture and optimizer, it reaches $99.6\,\%$. The dimension $n$ of the state space is therefore a genuine design parameter of planted-attractor classifiers: it has to be large enough for the trajectories of different classes to go around each other.

\subsection{A benchmark classification test: Fashion-MNIST}
\label{subsec:fmnist_bench}
\begin{figure}[H]
    \centering
    \includegraphics[width=\linewidth]{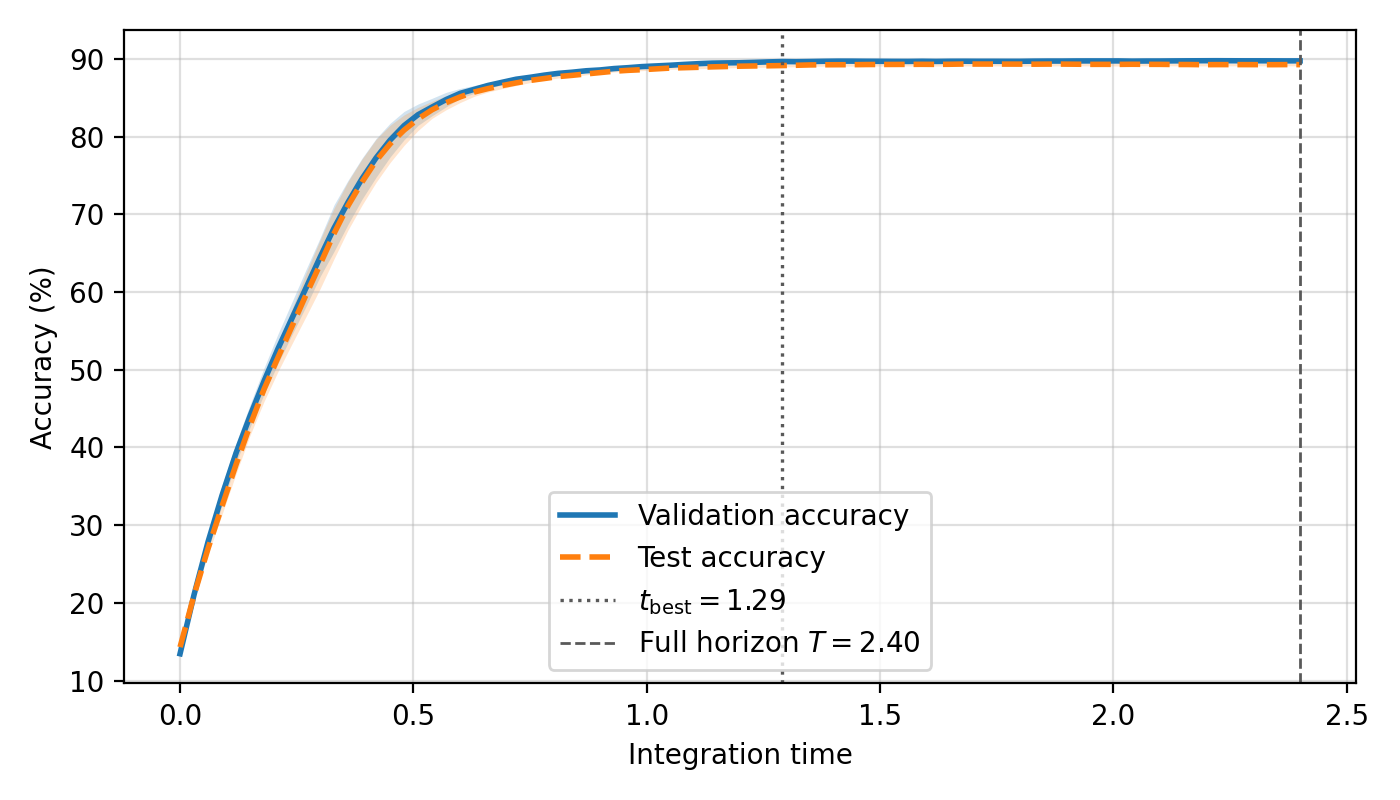}
    \caption{Accuracy of ConsNODEs on Fashion-MNIST as a function of the integration time (mean over the $5$ runs; the shaded bands show the standard deviation). The validation and test accuracy rapidly approach a plateau before the full training horizon $T$, showing that in practice the classifier can be evaluated with a shorter inference horizon $t_{\mathrm{best}}$ (dotted line), while essentially preserving the same accuracy.}
    \label{fig:fmnist_acc_time}
\end{figure}

{
In this section we consider the Fashion-MNIST dataset. We use neither convolutional encoders nor data augmentation. Each grayscale image is therefore flattened and supplied directly as an initial condition in the visible space,
\[
x(0)=x_0 \in \mathbb{R}^{784}.
\]
This experiment is meant to assess whether the QR-planted Neural-ODE classifier can remain competitive even in a bare setting where classification is performed directly from raw pixels.
We compare the QR-planted dynamical classifier (ConsNODEs) against a fully connected MLP with the same number of trainable parameters.
In the experiments reported here, ConsNODEs is instantiated in the visible space of dimension $n=784$, with hidden width $m=1000$, and the dynamics are integrated with an explicit Euler scheme using a fixed step size $\Delta t=0.03$ for a total of $80$ steps, corresponding to a terminal time $T=2.4$. ConsNODEs uses the ReLU activation and the block-coded equilibria of Sec.~\ref{subsec:dynamical_classification_principle}, with amplitude $a=0.8$. Of the $60{,}000$ training images, $6{,}000$ are held out for validation, and the pixels are standardized. ConsNODEs is trained for $30$ epochs with the Adam optimizer (learning rate $5\times10^{-4}$, batch size $256$) on the loss \eqref{eq:terminal_regression_loss}, to which a weak log-determinant penalty on the conditioning of the feature matrix $S$ is added (weight $10^{-3}$); the weights of the epoch with the best validation accuracy are retained. The MLP has three hidden layers of $699$, $698$ and $751$ ReLU units, which matches the number of parameters of ConsNODEs exactly; it is trained on the cross-entropy loss with Adam (learning rate $10^{-3}$, batch size $256$) for at most $30$ epochs, with early stopping on the validation accuracy (patience of $8$ epochs). The parameter-matched MLP is trained on the same train/validation/test split, so that both models are evaluated under identical data conditions. All the results are averaged over $5$ independent runs. The resulting test accuracy shows that the QR-planted dynamical classifier remains competitive with the static MLP baseline.
Table~\ref{tab:fashionmnist_compare} summarizes the quantitative comparison. Besides the final test accuracy, we report the number of trainable parameters, the estimated FLOPs for a single forward pass, the wall-clock training time per epoch, and the wall-clock inference time per sample. The latter is the end-to-end time of a pass over the test set and is dominated by the loading of the data, which is common to the two models; the forward pass alone is slower than that of the MLP by a factor of the order of the number of Euler steps, as indicated by the FLOPs. Since the parameter count is matched by construction, the table provides a direct comparison between a static feedforward architecture and a dynamical classifier with planted equilibria.
A distinctive feature of the planted-attractor formulation emerges when inspecting the accuracy as a function of integration time. After training, one can evaluate the dynamical classifier at intermediate times $t<T$, without retraining the model, by simply stopping the numerical integration at the desired time. The corresponding curve (accuracy vs.\ time) is reported in Fig.~\ref{fig:fmnist_acc_time}. As expected, the accuracy increases rapidly at early times and then saturates, reaching a plateau well before the terminal horizon $T$. This behavior has a natural dynamical interpretation. The classification rule follows the closest planted fixed point: once the trajectory has evolved sufficiently close to the corresponding equilibrium, pushing the integration further in time does not change the predicted label. Trajectories do not leave the basin of attraction, because the fixed point is stable. Additional integration steps refine the convergence towards the same fixed point, without producing a significant gain in classification accuracy.
This observation suggests a practical improvement at inference time. Instead of integrating the dynamics up to the training horizon $T$, one can define a shortened inference horizon $t_{\mathrm{best}}<T$ as the earliest integration time such that the achieved accuracy differs from the full-horizon accuracy by at most a given threshold. Here we use a threshold of $0.1$ percentage points on the validation accuracy averaged over the $5$ runs, which gives $t_{\mathrm{best}}=1.29$, i.e.\ $43$ Euler steps instead of $80$. In this way, the same trained model can be deployed with a reduced number of integration steps, thereby lowering both the effective inference time and the associated forward FLOPs, while preserving essentially the same predictive performance. Table~\ref{tab:fashionmnist_tbest} reports the comparison of Table~\ref{tab:fashionmnist_compare} when the integration of ConsNODEs is stopped at $t_{\mathrm{best}}$.}

\begin{table}[H]
\centering
\caption{Fashion-MNIST: comparison between a static MLP head and the QR-planted Neural-ODEs with the same number of parameters, with ConsNODEs integrated up to the training horizon $T=2.4$ ($80$ Euler steps).
Accuracy is reported as the mean $\pm$ standard deviation over $5$ runs.
FLOPs refer to a single forward pass with batch size $1$. Training times are wall-clock averages over the same runs. Inference times are end-to-end wall-clock times per sample of a pass over the test set (data loading, transfer to the GPU and forward pass, with batch size $256$).}

\label{tab:fashionmnist_compare}

\resizebox{\textwidth}{!}{
\begin{tabular}{lccccc}
\toprule
Model & Test Acc. (\%) & Params (total) &  FLOPs (forward) & Train time/epoch (s) & Infer time/sample (ms) \\
\midrule
MLP & $89.48 \pm 0.10$ & 1{,}569{,}784  & 3{,}135{,}252 & 1.62 & 0.033 \\
ConsNODEs & $89.28 \pm 0.16$ & 1{,}569{,}784  & 250{,}880{,}000 & 6.55 & 0.044 \\
\bottomrule
\end{tabular}
}
\end{table}

\begin{table}[H]
\centering
\caption{Fashion-MNIST: same comparison as in Table~\ref{tab:fashionmnist_compare}, with the integration of ConsNODEs stopped at the shortened inference horizon $t_{\mathrm{best}}=1.29$ ($43$ Euler steps). The trained models are those of Table~\ref{tab:fashionmnist_compare}: only the number of integration steps used at inference changes, and with it the test accuracy, the FLOPs and the inference time of ConsNODEs.}

\label{tab:fashionmnist_tbest}

\resizebox{\textwidth}{!}{
\begin{tabular}{lccccc}
\toprule
Model & Test Acc. (\%) & Params (total) &  FLOPs (forward) & Train time/epoch (s) & Infer time/sample (ms) \\
\midrule
MLP & $89.48 \pm 0.10$ & 1{,}569{,}784  & 3{,}135{,}252 & 1.62 & 0.033 \\
ConsNODEs & $89.16 \pm 0.18$ & 1{,}569{,}784  & 134{,}848{,}000 & 6.55 & 0.034 \\
\bottomrule
\end{tabular}
}
\end{table}

The curve of Fig.~\ref{fig:fmnist_acc_time} is an average over the test set, and could hide individual labels that keep changing along the trajectories. The per-sample diagnostics of Secs.~\ref{subsec:oscillation_classification} and~\ref{subsec:decision_times} do not rely on a visualization of the flow and can be computed in any dimension: applied to Fashion-MNIST in Appendix~\ref{sec:fmnist_per_sample}, they show that this is not the case. The classification oscillates only for a few percent of the inputs, mostly misclassified ones, and the switches take place close to the decision boundary; the labels are fixed early, and stopping the integration at $t_{\mathrm{best}}$ leaves them essentially unchanged.

{The QR-planted Neural-ODE does not improve upon the MLP in terms of raw accuracy, training cost, or forward FLOPs on this benchmark. This was not, however, the primary purpose of the comparison. Rather, the aim was to show that a classifier built as a structured dynamical system with planted class equilibria can attain essentially the same predictive performance as a standard parameter-matched MLP. From this perspective, the relevance of the result is conceptual rather than purely computational: the decision mechanism is no longer entrusted to a static feedforward map. Instead, it follows a dynamical process that drives the inputs towards class-dependent equilibria. The resulting classifier, while achieving essentially the same accuracy, lends itself to the dynamical analysis discussed in Sec.~\ref{subsec:interpretability}: an example is the study of the stability of the planted equilibria reported in Sec.~\ref{subsec:unstable_to_stable}, another one the per-sample diagnostics of Appendix~\ref{sec:fmnist_per_sample}.}

\subsection{Emergence of stability during training}
\label{subsec:unstable_to_stable}

We use the same model and training loss as in the Fashion-MNIST experiment above ($n=784$, $m=1000$, ReLU activation, explicit Euler scheme with $\Delta t=0.03$ and $80$ steps, terminal loss \eqref{eq:terminal_regression_loss}, Adam optimizer), which also includes a weak log-determinant conditioning penalty on the feature matrix $S$ (weight $10^{-3}$). No term of the objective involves the Jacobian, and no stability regularizer is used. The differences with respect to the main experiment are the initialization of the parameters, described below, the settings of the optimizer (learning rate $10^{-3}$ for $W$ and $b_1$ and $10^{-4}$ for $A_2$ and $b_2$, with weight decay $10^{-4}$ and $5\times10^{-4}$, respectively), and the length of the training ($80$ epochs instead of $30$); we report the weights at the end of the training, without selecting the epoch on the validation accuracy. The Jacobians are evaluated at the nominal prototypes $\bar x^{(\ell)}$ through Eq.~\eqref{eq:jacobian_param}, and their $n=784$ eigenvalues are computed by dense eigendecomposition, for each of the $C=10$ equilibria ($7840$ eigenvalues in total).

\begin{figure}[H]
    \centering
    \includegraphics[width=\linewidth]{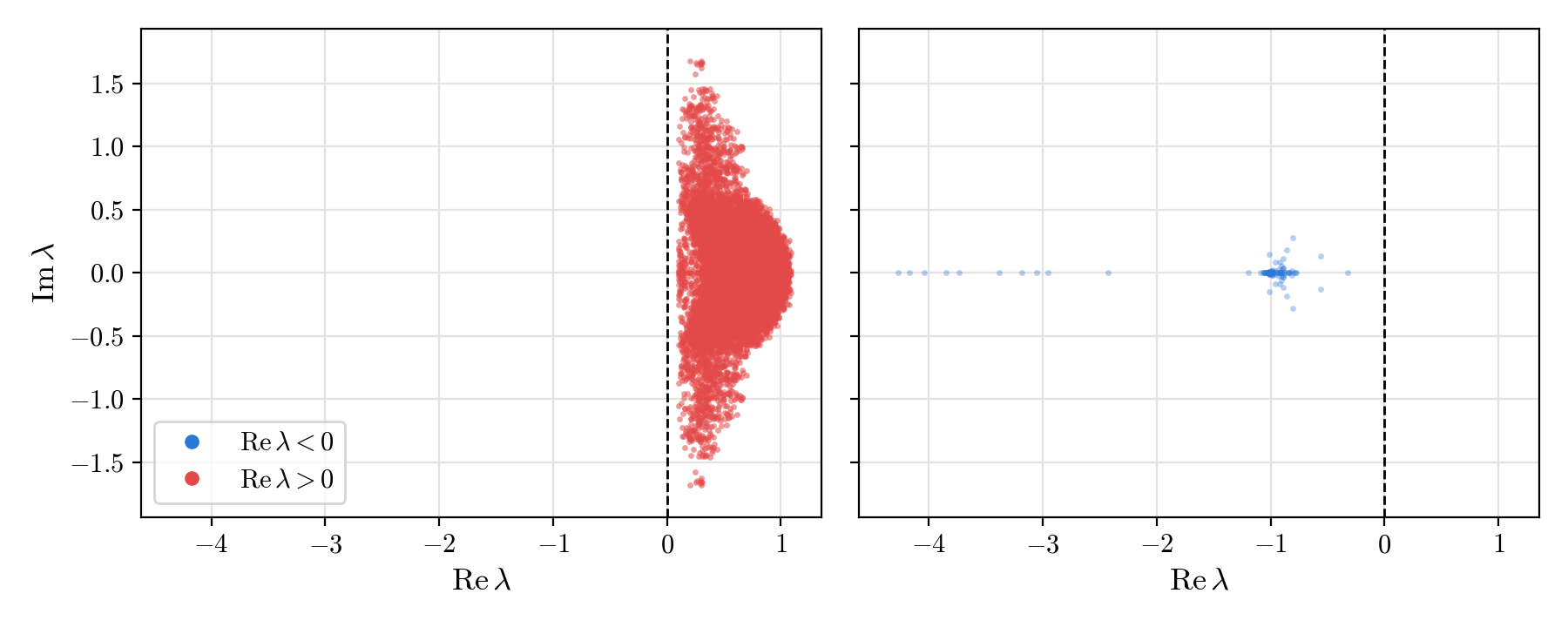}
    \caption{Fashion-MNIST: spectrum of the Jacobian $J^{(\ell)}$, Eq.~\eqref{eq:jacobian_param}, at the $C=10$ planted fixed points of ConsNODEs. The $n=784$ eigenvalues of each of the $10$ Jacobians are overlaid in the complex plane ($10\times784=7840$ points per panel; horizontal axis $\operatorname{Re}\lambda$, vertical axis $\operatorname{Im}\lambda$, same axes in both panels). Red (blue) points have $\operatorname{Re}\lambda>0$ ($\operatorname{Re}\lambda<0$). Left: at initialization, engineered so that all the eigenvalues are unstable ($0.1\le\operatorname{Re}\lambda\le1.09$). Right: after $80$ epochs of training with the classification loss alone, all the eigenvalues have negative real part ($\max_{\ell}\alpha_\ell=-0.32$). Single run (seed $0$); test accuracy $87.5\%$.}
    \label{fig:jacobian_stability}
\end{figure}

\emph{Result.} The initial Jacobian is unstable in every direction at every equilibrium: all the $7840$ eigenvalues lie in the right half-plane, with $0.1\le\operatorname{Re}\lambda\le1.09$ (left panel of Fig.~\ref{fig:jacobian_stability}). After training the picture is reversed (right panel): all the $7840$ eigenvalues have negative real part, the spectral abscissa of each equilibrium is negative ($\max_\ell\alpha_\ell=-0.32$), and most of the eigenvalues collapse into a compact cluster around $-1$, the value of the pure leak term $-x$. The trained classifier reaches a test accuracy of $87.5\%$.

The stability of the planted equilibria need therefore not be imposed: it is learned. No term of the objective rewards a negative spectral abscissa, and its emergence is a consequence of minimizing the terminal loss \eqref{eq:terminal_regression_loss}. Heuristically, near an unstable equilibrium a deviation grows as $e^{\alpha_\ell t}$ along the unstable eigendirections, so that a trajectory approaching $\bar x^{(\ell)}$ is expelled rather than captured, and the distance $\|x(T)-\bar x^{(\ell)}\|_2$ cannot be made small. To reduce the loss the system has to turn each planted equilibrium into an attractor, which is precisely the property that the classification rule \eqref{eq:nearest_attractor_rule} relies on.

\subsection{A more challenging test: the case of CIFAR-10}
\label{subsec:models_compared}

Given an input image $u\in\mathbb R^{3\times 32\times 32}$, a convolutional feature extractor $\phi$ produces a compact
representation $x_0 := \phi(u)\in\mathbb R^n$.

In all experiments we fix $n=320$ and $m=512$. The CNN is shared \emph{identically} by the two models, so that differences in
performance can be attributed to the classification head. We use CIFAR-10 without data augmentation.

The feature extractor $\phi$ consists of three convolutional blocks with $64$, $128$ and $256$ channels, each made of two $3\times3$ convolutions followed by batch normalization and a ReLU activation, and closed by a $2\times2$ max-pooling; the resulting $4096$ features are mapped to $\mathbb R^n$ by a fully connected layer with $512$ ReLU units (with dropout probability $0.2$) followed by a linear layer. It is trained end-to-end together with the head, starting from the same initial weights for the two models. The ConsNODEs head uses the sigmoid activation and the block-coded equilibria of Sec.~\ref{subsec:dynamical_classification_principle} with amplitude $a=3$, and is integrated with the explicit Euler scheme with $\Delta t=0.03$ for $80$ steps ($T=2.4$). The MLP head applies a layer normalization and a ReLU to $x_0$ and has two hidden layers of $272$ and $850$ ReLU units, with dropout probability $0.2$, which gives exactly the same number of parameters as the ConsNODEs head.

We train each model for the same number of epochs using identical data processing, optimizer settings, and batch size.
Specifically, both models are trained for $80$ epochs with the Adam optimizer (learning rate $10^{-3}$, no weight decay) and batch size $256$, with neither a learning-rate schedule nor early stopping, and the model obtained at the last epoch is evaluated; ConsNODEs is trained on the loss \eqref{eq:terminal_regression_loss} and the MLP on the cross-entropy loss. Of the $50{,}000$ training images, $5{,}000$ are held out for validation.
To quantify run-to-run variability, we repeat training for $5$ independent random seeds and report the mean and standard
deviation of the final test accuracy. In addition to accuracy, we report:
(i) the number of trainable parameters,
(ii) forward-pass FLOPs for a single input,
(iii) wall-clock time per training epoch, and
(iv) wall-clock inference time per sample.

\begin{figure}[H]
    \centering
     \includegraphics[width=0.90\linewidth]{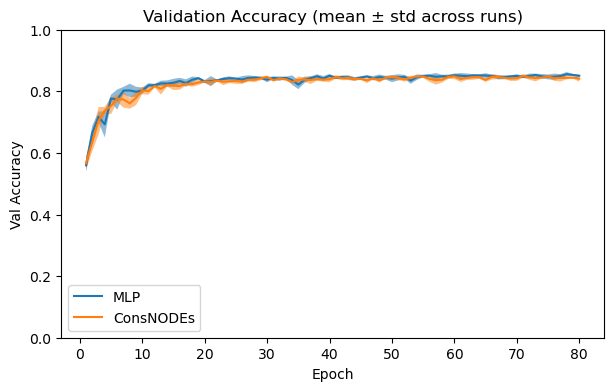}
    \caption{CIFAR-10 accuracy learning curves (mean $\pm$ standard deviation over $5$ runs) comparing
    the parameter-matched MLP head and the QR-planted dynamical head (ConsNODEs).}
    \label{fig:cifar10_learning_curves}
\end{figure}

Figure~\ref{fig:cifar10_learning_curves} reports the learning curves (mean $\pm$ standard deviation over the $5$ runs)
for the two heads.

Table~\ref{tab:cifar10_summary} summarizes the final results. Both models have the same number of trainable parameters
(3.74M total, 0.33M in the head) and achieve comparable test accuracy. The dynamical head incurs additional
computation due to ODE integration and the QR/solve step required to enforce \eqref{eq:equilibrium_constraints},
resulting in higher FLOPs and longer wall-clock time.

\begin{table}[H]
\centering
\caption{CIFAR-10 comparison between a parameter-matched static MLP head and the QR-planted dynamical head (ConsNODEs).
Accuracy is reported as mean $\pm$ standard deviation over $5$ runs.
FLOPs refer to a single forward pass with batch size $1$. Training and inference times are wall-clock averages over the same runs.}
\label{tab:cifar10_summary}

\resizebox{\textwidth}{!}{
\begin{tabular}{lcccccc}
\toprule
Model & Test Acc. (\%) & Params (total) & Params (head) & FLOPs (forward) & Train time/epoch (s) & Infer time/sample (ms) \\
\midrule
MLP & $84.69 \pm 0.46$ & 3{,}737{,}536 & 328{,}512 & 310.7M & 3.23 & 0.0289 \\
ConsNODEs & $84.39 \pm 0.62$ & 3{,}737{,}536 & 328{,}512 & 375.9M & 5.77 & 0.0406 \\
\bottomrule
\end{tabular}
}
\end{table}

From an accuracy standpoint, the two heads are essentially on par at this scale: the $0.30$ percentage point difference
in mean test accuracy is within one standard deviation. The main cost of enforcing exact planted equilibria through the
QR parametrization is computational. Relative to the MLP head, ConsNODEs increases forward FLOPs by approximately
$21\%$ (310.7M $\to$ 375.9M) and increases the average training time per epoch by a factor of $\approx 1.8$
(3.23\,s $\to$ 5.77\,s), while inference latency rises from 0.0289\,ms to 0.0406\,ms per sample. These overheads reflect the
explicit time integration and the additional linear-algebra operations required to maintain the fixed-point constraints
\eqref{eq:equilibrium_constraints} throughout training.

The relevance of the proposed dynamical head should therefore not be assessed solely in terms of raw computational efficiency or marginal accuracy gains over a standard MLP. Rather, the point of the experiment is to show that a classifier whose decision mechanism is implemented by a continuous-time dynamical system with hard-planted class equilibria can achieve essentially the same predictive performance as a parameter-matched feedforward head. Moreover, by exploiting the saturation of accuracy along the integration trajectory, the inference horizon can be shortened after training, reducing the effective cost of the dynamical classifier and bringing its latency close to that of the static MLP baseline. In this sense, the proposed model trades a moderate computational overhead for a qualitatively different classification mechanism: inputs are transported by a learned vector field toward class-dependent equilibria, so that the classifier can be investigated with the tools of dynamical systems theory, in the sense of the dynamical interpretability discussed in Sec.~\ref{subsec:interpretability}.

\begin{figure}[H]
    \centering
    \includegraphics[width=0.7\linewidth]{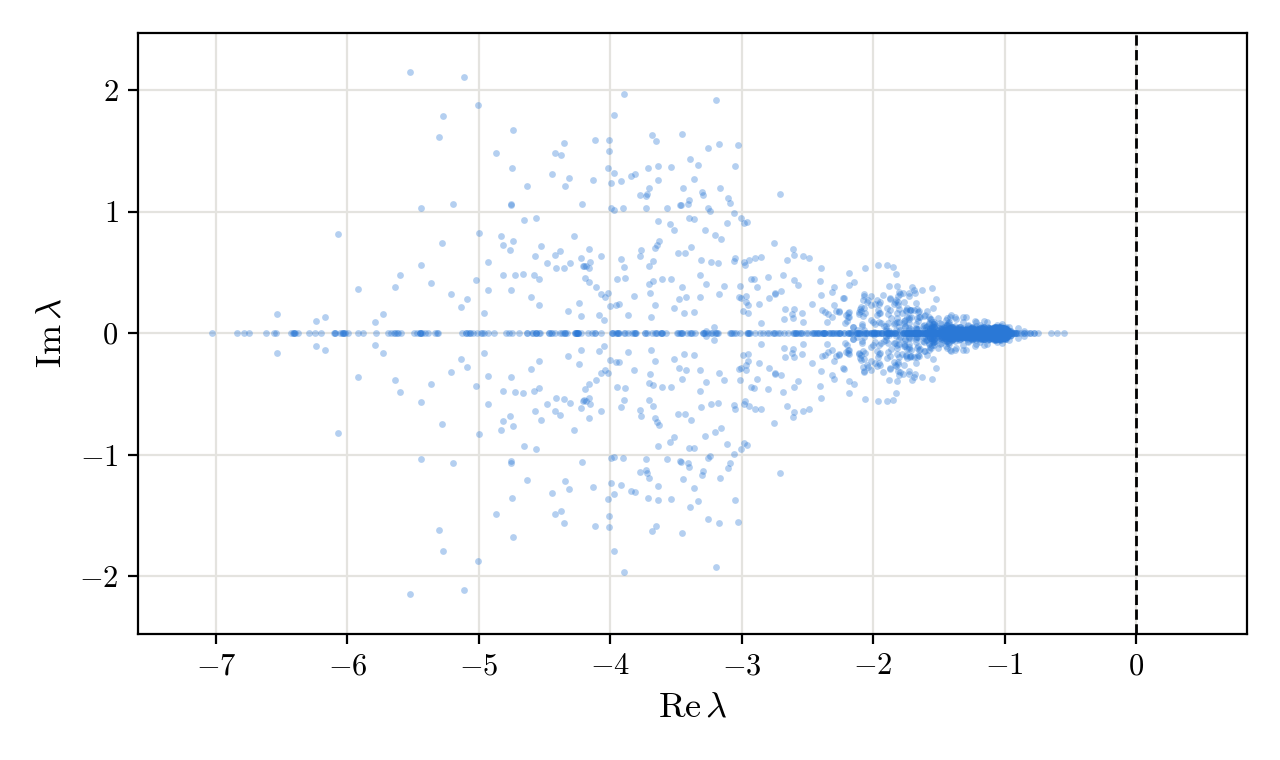}
    \caption{CIFAR-10: spectrum of the Jacobian $J^{(\ell)}$, Eq.~\eqref{eq:jacobian_planted}, at the $C=10$ planted equilibria of the ConsNODEs head, for the five trained models of Table~\ref{tab:cifar10_summary}. The $n=320$ eigenvalues of each Jacobian are overlaid in the complex plane ($5\times10\times320=16{,}000$ points; horizontal axis $\operatorname{Re}\lambda$, vertical axis $\operatorname{Im}\lambda$). The dashed line marks $\operatorname{Re}\lambda=0$. All the eigenvalues have negative real part ($\max_\ell\alpha_\ell=-0.55$ over the five runs).}
    \label{fig:cifar10_jacobian_spectrum}
\end{figure}

As an example of such an analysis, we report the spectrum of the Jacobian at the planted equilibria, as done for Fashion-MNIST in Sec.~\ref{subsec:unstable_to_stable}. The dynamical head acts in the feature space $\mathbb R^{n}$, with $n=320$: the Jacobian $J^{(\ell)}$ of Eq.~\eqref{eq:jacobian_planted} involves only the parameters of the head, while the feature extractor $\phi$ sets the initial condition. Figure~\ref{fig:cifar10_jacobian_spectrum} shows the eigenvalues of the $C=10$ Jacobians of the five trained models of Table~\ref{tab:cifar10_summary}, i.e.\ $5\times10\times320=16{,}000$ eigenvalues. All of them have negative real part, and the spectral abscissa of the $50$ equilibria lies in the interval $-0.99\le\alpha_\ell\le-0.55$: every planted point is an attractor of the learned flow, in each of the five runs. The same holds for the explicit Euler map that is iterated in practice, since $|1+\Delta t\,\lambda|\le0.983$ for all the eigenvalues. As for Fashion-MNIST, no term of the loss involves the Jacobian. Here, moreover, the initialization is the standard one of the experiment and it is not stable: before training all the ten equilibria are weakly unstable, with $0.01\le\alpha_\ell\le0.04$ and about $1\,\%$ of the eigenvalues in the right half-plane ($161$ out of $16{,}000$ over the five runs). Stability is thus acquired during training also in this case, without any ad hoc initialization.

\section{Conclusion}

In this work, we bridged a gap between two independent lines of research. On the one hand, dynamical-system-based classifiers with planted attractors provide a natural approach to classification, which is interpretable in the dynamical sense discussed in Sec.~\ref{subsec:interpretability}. Classes are encoded as asymptotically stable states of the examined dynamics. The ensuing decision-making process can be analyzed via standard methodologies, typical of the theory of dynamical systems. These include, but are not limited to, the concept of basin of attraction, local stability analysis, Lyapunov exponents, and the evolution of probability densities. However, previous attractor-based constructions did not come with universality guarantees. Their expressive power on generic classification tasks therefore remained unclear. On the other hand, Neural-ODEs are universal approximators of vector fields, but in standard formulations they are not directly designed to enforce, \emph{a priori}, a prescribed set of stable fixed points. Classification is therefore typically performed by resorting to an external readout head. As demonstrated in \cite{pacificoNEURIPS}, it is possible to embed a finite set of prescribed fixed points into a Neural-ODE architecture while maintaining universal approximation within the class of vector fields compatible with those constraints. Crucially, the introduction of these fixed points does not compromise the model's expressivity; the architecture remains sufficiently flexible to approximate arbitrary residual dynamics, provided they satisfy the required fixed-point conditions. Furthermore, by virtue of its design, gradient-based training is structurally confined to this constrained hypothesis class, ensuring that the desired dynamical properties are preserved throughout the optimization process.

 In this work, Neural-ODEs equipped with a curated collection of equilibrium points have been successfully employed for classification tasks. The planted attractors serve as indicators for the target classes, while the velocity field, leveraging the universal approximation capabilities of the architecture, sculpts the dynamical landscape. This process defines the basins of attraction of the trained model, effectively directing each input (provided as an initial condition) toward its corresponding destination target.

Finally, we showed that the proposed framework admits, in a suitable limit, an equivalent interpretation in terms of a coupled multi-population dynamical system, in the spirit of the Funahashi--Nakamura construction \cite{FUNAHASHI1993801}. This provides a structural reading of the classifier: the trainable matrices $A_1$ and $A_2$ are the weighted couplings between the visible and the hidden units, and the stability of the planted equilibria is controlled by the gain of the loop that they form. Beyond this, the reading has not been exploited in the present work. Exploring this connection further, and exploiting it to derive new insights on learning and classification from a dynamical-systems and neuroscientific perspective, represents an interesting avenue for future studies.

\newpage
\appendix
\section{Sensitivity to the distance between the planted equilibria}
\label{sec:attractor_distance}

The planting construction of Sec.~\ref{subsec:node_setup} does not prescribe a preferred
location for the class equilibria, beyond requiring the feature matrix
$S$ to have full column rank. In practice, however, the distance between
the prescribed equilibria cannot be chosen arbitrarily. Two opposite
regimes are potentially problematic: equilibria that are too close make
the planting constraints nearly linearly dependent, whereas equilibria
that are very far from the data require increasingly strong finite-time
transport. We discuss these two limits separately.

Let us begin by considering the setting where equilibria are too close.
Let
\[
d_{\ell k}
:=
\left\|
\bar{x}^{(\ell)}-\bar{x}^{(k)}
\right\|_2 ,
\]
be the distance between two planted equilibria and recall that
\[
s_\ell
=
f\!\left(A_2\bar{x}^{(\ell)}+b_2\right),
\qquad
S=[s_1,\ldots,s_C].
\]
If the activation $f$ is $L_f$-Lipschitz, then, for every
$\ell\neq k$, the smallest singular value of $S$ satisfies
\begin{equation}
\sigma_{\min}(S)
\leq
\frac{\|s_\ell-s_k\|_2}{\sqrt{2}}
\leq
\frac{L_f\|A_2\|_2}{\sqrt{2}}\,d_{\ell k}.
\label{eq:smin_distance}
\end{equation}
Indeed, the first inequality follows by evaluating $S$ along the unit
direction $(e_\ell-e_k)/\sqrt{2}$, while the second follows directly
from the Lipschitz continuity of $f$.

Equation~\eqref{eq:smin_distance} shows that, at fixed $A_2$, the
smallest singular value is upper bounded by a quantity proportional to
the separation of the closest pair of equilibria. Hence,
\[
d_{\ell k}\rightarrow 0
\qquad\Longrightarrow\qquad
\sigma_{\min}(S)\rightarrow 0.
\]
In the limiting case
$\bar{x}^{(\ell)}=\bar{x}^{(k)}$, the two feature vectors coincide,
$s_\ell=s_k$, so that two columns of $S$ are identical and
$\operatorname{rank}(S)<C$. Thus, when two fixed points become
too close, the corresponding planted constraints become increasingly
difficult to distinguish.

Consider now the case of equilibria that are far apart.
A different limitation arises when the target equilibria are moved far
away from the region occupied by the data while the integration horizon
$T$ is kept fixed. For every trajectory,
\[
x(T)-x_0
=
\int_0^T F_\theta(x(t))\,dt,
\]
and therefore
\begin{equation}
\max_{0\leq t\leq T}
\|F_\theta(x(t))\|_2
\geq
\frac{\|x(T)-x_0\|_2}{T}.
\label{eq:transport_bound}
\end{equation}
If the trajectory is required to end within an error
$\varepsilon_\ell=
\|x(T)-\bar{x}^{(\ell)}\|_2$
from its prescribed equilibrium, the triangle inequality gives
\begin{equation}
\max_{0\leq t\leq T}
\|F_\theta(x(t))\|_2
\geq
\frac{
\|\bar{x}^{(\ell)}-x_0\|_2-\varepsilon_\ell
}{T}.
\label{eq:transport_bound_target}
\end{equation}
Thus, independently of the activation function, moving the targets
arbitrarily far from the data at fixed $T$ requires the magnitude of
the vector field along the corresponding trajectories to become
arbitrarily large. In other words, distant targets impose an
increasingly demanding finite-time transport problem.

For bounded activations (as for sigmoidal nonlinear filters) this requirement has a further consequence at
the level of the network parameters. Subtracting the fixed-point
conditions for two equilibria gives
\[
A_1(s_\ell-s_k)
=
\bar{x}^{(\ell)}-\bar{x}^{(k)}.
\]
Let $m$ denote the number of hidden units, i.e.\ the dimension of each
feature vector $s_\ell\in\mathbb{R}^m$, and define
\[
C_f
:=
\sup_{u,v\in\mathbb{R}}
|f(u)-f(v)| ,
\]
which is finite whenever $f$ is bounded. Since $f$ acts
component-wise,
\[
\|s_\ell-s_k\|_2
\leq C_f\sqrt{m},
\]
and hence
\begin{equation}
\|A_1\|_2
\geq
\frac{d_{\ell k}}{C_f\sqrt{m}}.
\label{eq:A1_far}
\end{equation}
For the logistic sigmoid, for instance, $C_f=1$. Therefore, at fixed
hidden width $m$, increasing the mutual distance between the planted
equilibria forces the norm of the output matrix $A_1$ to grow at least
linearly. Although large weights do not by themselves imply dynamical
instability, they are generally undesirable from an optimization and
numerical perspective, since they increase parameter sensitivity and
can lead to poorly conditioned or unstable training dynamics.

The previous argument does not apply directly to the ReLU activation,
since ReLU is unbounded. In this case the increase in the target scale
can be absorbed by the hidden representation itself rather than by
$A_1$. To see this, consider a family of equilibria whose overall scale
is controlled by a parameter $v$,
\[
\bar{x}^{(\ell)}(v)=v q_\ell.
\]
For ReLU, one gets
\[
s_\ell(v)
=
\operatorname{ReLU}
\left(vA_2q_\ell+b_2\right)
=
v\,
\operatorname{ReLU}
\left(A_2q_\ell+\frac{b_2}{v}\right).
\]
We have thus two limiting possibilities: (i) $s_\ell(v)$ stays finite, if
$\|A_2\|_2$ becomes smaller when training the model for larger $v$; (ii)
$s_\ell(v)$ diverges, if $\|A_2\|_2$ stays finite when training the model for
larger $v$. Since $Y$ (the right-hand side of the planting condition
$A_1S=Y$) grows as $v$ (and thus diverges) we cannot come to a definite
conclusion on the divergence of $\|A_1\|_2$, as opposed to the case of a
bounded activation function.

\begin{figure}[htbp]
    \centering
    \begin{subfigure}{0.48\linewidth}
        \centering
        \includegraphics[width=\linewidth]{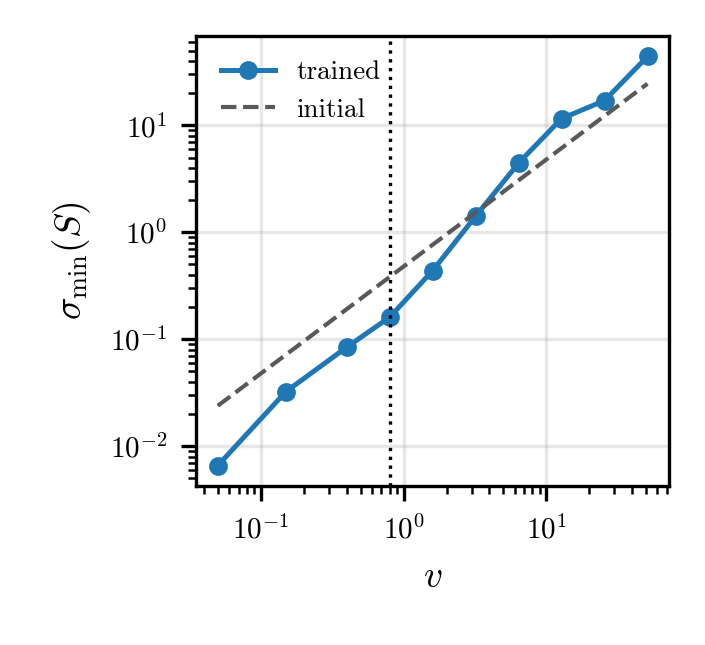}
        \caption{}
        \label{fig:attractor_distance_fmnist_smin}
    \end{subfigure}
    \hfill
    \begin{subfigure}{0.48\linewidth}
        \centering
        \includegraphics[width=\linewidth]{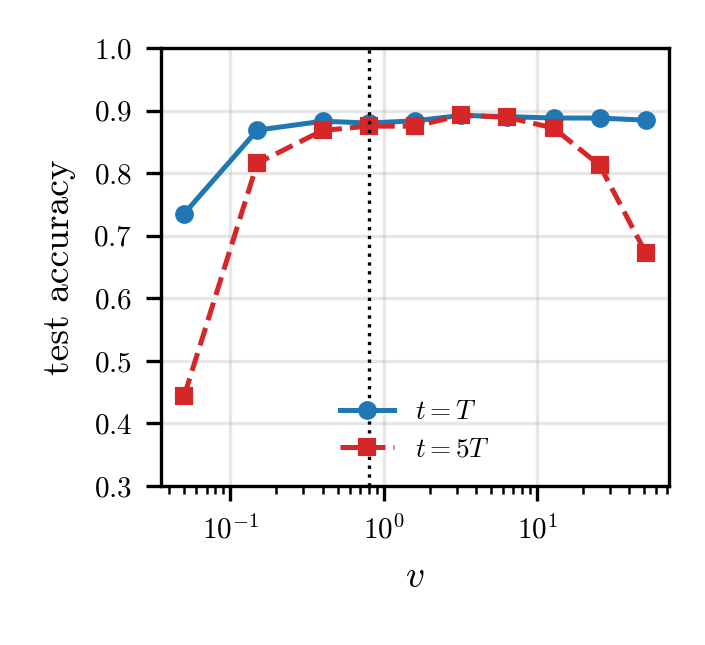}
        \caption{}
        \label{fig:attractor_distance_fmnist_acc}
    \end{subfigure}

    \vspace{0.5em}
    \begin{subfigure}{0.48\linewidth}
        \centering
        \includegraphics[width=\linewidth]{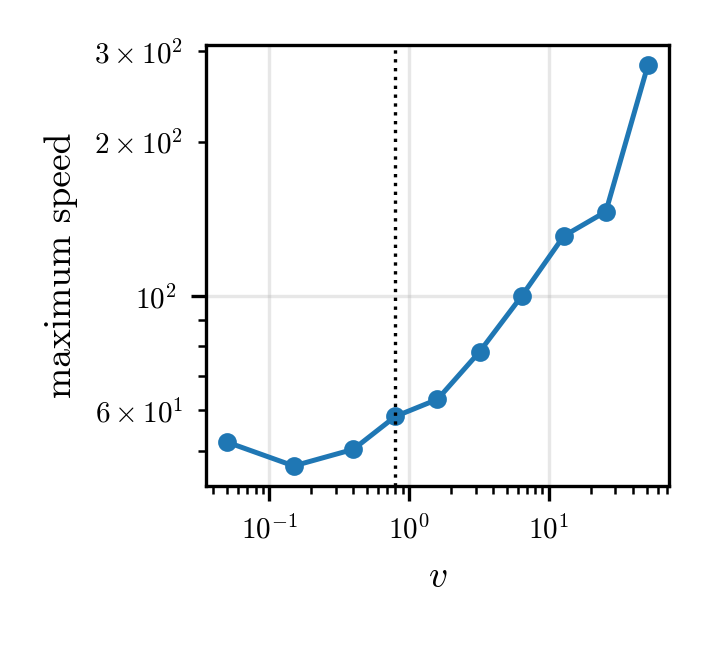}
        \caption{}
        \label{fig:attractor_distance_fmnist_speed}
    \end{subfigure}
    \hfill
    \begin{subfigure}{0.48\linewidth}
        \centering
        \includegraphics[width=\linewidth]{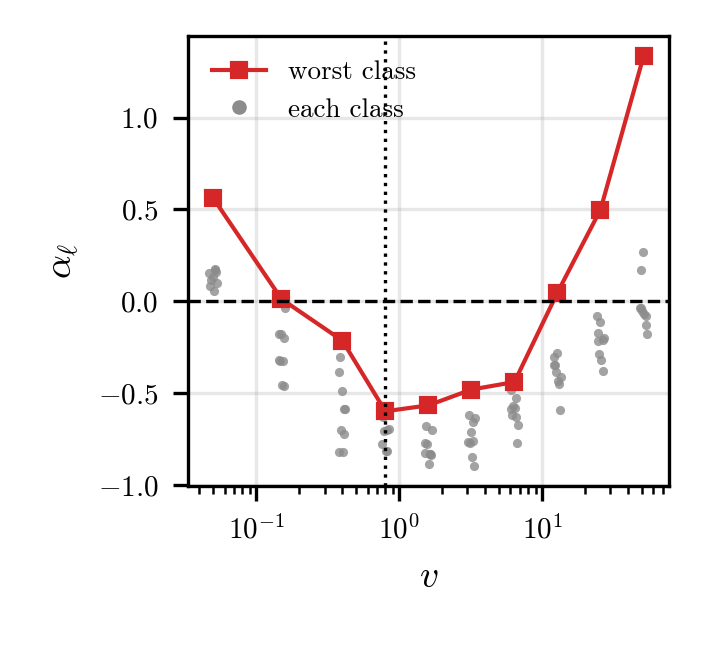}
        \caption{}
        \label{fig:attractor_distance_fmnist_alpha}
    \end{subfigure}
\caption{
    Sensitivity of the Fashion-MNIST classifier to the distance between the planted equilibria.
    The $C=10$ equilibria are generated with the same block construction used in Sec.~\ref{subsec:fmnist_bench},
    with common amplitude $v$, so that their mutual distance scales linearly with $v$.
    All other architectural and training choices are kept unchanged.
    (a) Smallest singular value $\sigma_{\min}(S)$ of the feature matrix after training
    (solid line) and at initialization (dashed line). For small $v$, the planted features
    become increasingly difficult to distinguish and $S$ approaches rank deficiency.
    (b) Test accuracy of the nearest-equilibrium decoder evaluated at the training horizon
    $T$ and after a longer integration time $5T$.
    (c) Median, over the test trajectories, of the maximum speed
    $\max_{0\leq t\leq T}\|F_\theta(x(t))\|_2$.
    The increase at large $v$ reflects the finite-time transport requirement
    in Eq.~\eqref{eq:transport_bound_target}.
    (d) Spectral abscissa
    $\alpha_\ell=\max_i \operatorname{Re}\lambda_i(J_\theta(\bar{x}^{(\ell)}))$
    of each planted equilibrium.
    Gray points denote the individual equilibria, while the highlighted curve reports
    the least stable equilibrium.
    The horizontal dashed line $\alpha_\ell=0$ separates asymptotically stable
    from unstable fixed points.
    The vertical dotted line marks the value used in the reference experiment of
    Sec.~\ref{subsec:fmnist_bench}.
    }
\label{fig:attractor_distance_fmnist}
\end{figure}

What remains unavoidable, also in the case of a ReLU activation, is the transport requirement
\eqref{eq:transport_bound_target}. Even when the scaling is absorbed by
the ReLU features rather than by the output weights, trajectories must
still travel increasingly large distances within the same finite time
horizon.
We illustrate these two limiting regimes using the Fashion-MNIST
classifier of Sec.~\ref{subsec:fmnist_bench}, for which the Neural-ODE employs a ReLU
activation. The $C=10$ class equilibria are defined through the same
block construction used in the main experiment, but their common
amplitude $v$ is varied over a wide range while all other architectural
and training choices are kept unchanged. The results are summarized in
Fig.~\ref{fig:attractor_distance_fmnist}.

For small $v$, the numerical results are consistent with
Eq.~\eqref{eq:smin_distance}: the smallest singular value of $S$
decreases rapidly as the equilibria approach each other
(Fig.~\ref{fig:attractor_distance_fmnist}a), and the
condition number increases at the smallest separations. This is
accompanied by a deterioration of classification accuracy
(Fig.~\ref{fig:attractor_distance_fmnist}b), indicating
that the regime of nearly coincident class prototypes is unfavorable
both algebraically and from the viewpoint of the nearest-equilibrium
decoder.

The large-$v$ regime is qualitatively different. In the trained models
the norm of $A_1$ does not grow with the target scale; since $Y$ grows as
$v$, the planting condition $A_1S=Y$ then forces the hidden features to
become increasingly large, i.e.\ the trained models realize the second of
the two possibilities discussed above.
At the same time, the trajectories become substantially longer and
their maximum speed increases (Fig.~\ref{fig:attractor_distance_fmnist}c),
in agreement with the finite-time
transport bound~\eqref{eq:transport_bound_target}. Thus, for ReLU,
distant targets do not necessarily manifest themselves through
diverging output weights, but rather through increasingly large feature
amplitudes and increasingly fast dynamics.

We also observe a loss of local stability of some of the planted fixed
points at the largest values of $v$. Denoting by
\[
\alpha_\ell
=
\max_i
\operatorname{Re}\lambda_i
\left(
J_\theta(\bar{x}^{(\ell)})
\right)
\]
the spectral abscissa of the Jacobian at equilibrium $\ell$, some
trained models exhibit $\alpha_\ell>0$ for sufficiently distant
targets (Fig.~\ref{fig:attractor_distance_fmnist}d). This instability is \emph{not} a theoretical consequence of
moving the equilibria far away in the ReLU model: the planting
construction fixes the location of the equilibria but does not enforce
their stability. Rather, it is an empirical property of the solutions
selected by optimization in this extreme transport regime. Its effect
becomes evident when the system is integrated beyond the training
horizon: trajectories may pass close enough to an unstable planted
equilibrium to be classified correctly at time $T$, but are subsequently
repelled from it, leading to a degradation of the long-time
nearest-equilibrium accuracy.

Overall, the two extreme regimes point to a simple trade-off in the
placement of the class prototypes. Equilibria should be sufficiently
separated so that the corresponding planting features remain
distinguishable, but they should not be placed so far from the data
that classification requires excessively strong finite-time transport.
In our Fashion-MNIST experiment, a broad intermediate range of
equilibrium separations satisfies both requirements, suggesting that
placing the prototypes on a scale comparable with characteristic
distances in the data provides a useful practical initialization
heuristic, rather than a universal prescription.

\section{Per-sample diagnostics of the classification on Fashion-MNIST}
\label{sec:fmnist_per_sample}

We apply here to Fashion-MNIST the diagnostics introduced for the two spirals in Secs.~\ref{subsec:oscillation_classification} and~\ref{subsec:decision_times}, on the five models of Table~\ref{tab:fashionmnist_compare}, integrating the flow up to $t=5T$. The numbers below are averages over the five models and refer to the test set; the training set gives the same picture.

A first difference with the spirals concerns the initial condition. The standardized images lie at almost the same distance from all the planted equilibria, so that the nearest-equilibrium rule applied at $t=0$ is close to chance: the label is assigned by the flow, and most trajectories cross the boundary $\tilde m_i=0$ once, early in the integration. With ten classes the label can also switch between two wrong classes without crossing this boundary; such switches are frequent in the early transient, and are included in the decision times $\tau_i$.

\begin{figure}[htbp]
    \centering
    \begin{subfigure}{0.48\linewidth}
        \centering
        \includegraphics[width=\linewidth]{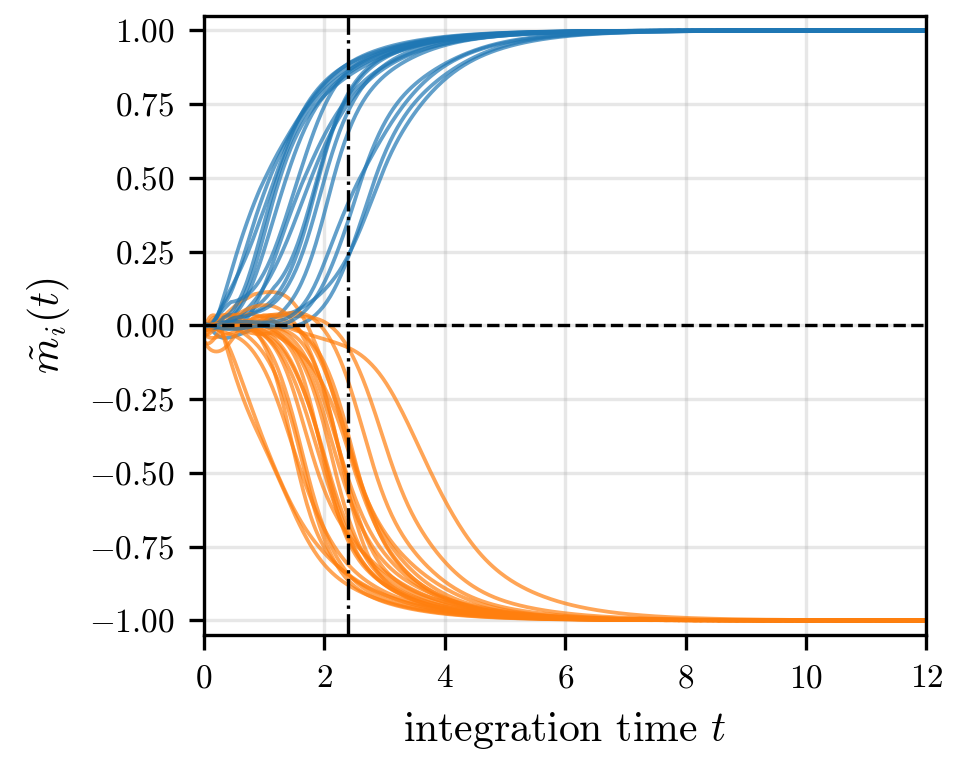}
        \caption{Up to $t=5T$}
        \label{fig:fmnist_oscillation_full}
    \end{subfigure}
    \hfill
    \begin{subfigure}{0.48\linewidth}
        \centering
        \includegraphics[width=\linewidth]{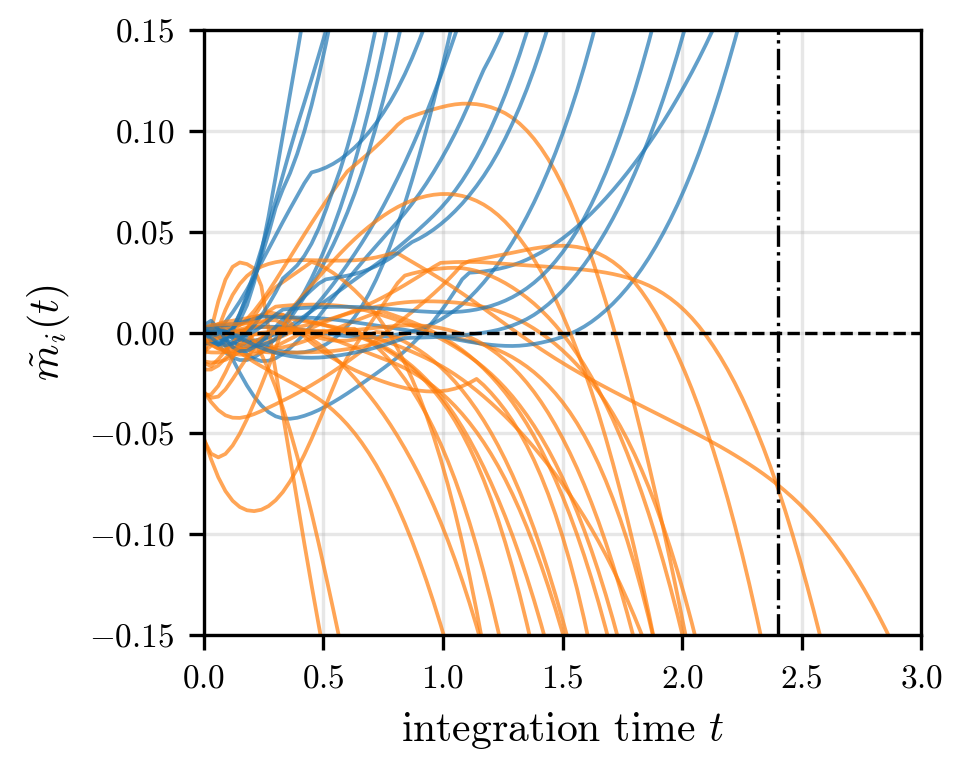}
        \caption{Zoom on $|\tilde m_i|\le0.15$}
        \label{fig:fmnist_oscillation_zoom}
    \end{subfigure}
    \caption{Oscillation of the classification on Fashion-MNIST (test set). Normalized margin $\tilde m_i(t)$, Eq.~\eqref{eq:mtilde}, of $40$ test inputs whose classification oscillates ($\tilde m_i$ crosses zero at least twice along $[0,T]$), drawn at random among those of one of the five models ($3.9\,\%$ of its test set). Blue: correctly classified at $T$; orange: misclassified at $T$. The horizontal dashed line marks the boundary $\tilde m_i=0$ between the correct and the wrong classes, and the dash-dotted vertical line the training horizon $T=2.4$.
    (a) Integration up to $t=5T=12$: every trajectory converges to the equilibrium of the correct class ($\tilde m_i\to1$) or to that of a wrong one ($\tilde m_i\to-1$).
    (b) The same trajectories for $t\le3$ and $|\tilde m_i|\le0.15$: the switches of the label take place while $|\tilde m_i|$ is small.}
    \label{fig:fmnist_oscillation}
\end{figure}

The classification oscillates for $3.5\,\%$ of the inputs, against $8.4\,\%$ for the spirals with one turn. These are mostly inputs that end up misclassified, and belong mainly to the upper-body garments (coat, T-shirt/top, shirt and pullover), the classes most easily confused with one another. The oscillations also differ in nature from those of the spirals. There a trajectory enters deep into the region of a wrong class before coming back (Fig.~\ref{fig:oscillation_plateau_spirals}); here the label switches while the trajectory is still almost equidistant from two competing equilibria (the margin $|\tilde m_i|$ reached before the last switch is typically a few hundredths, against about $0.7$ for the spirals), and once the trajectory has left this region the label no longer changes (Fig.~\ref{fig:fmnist_oscillation}).

\begin{figure}[htbp]
    \centering
    \begin{subfigure}{0.48\linewidth}
        \centering
        \includegraphics[width=\linewidth]{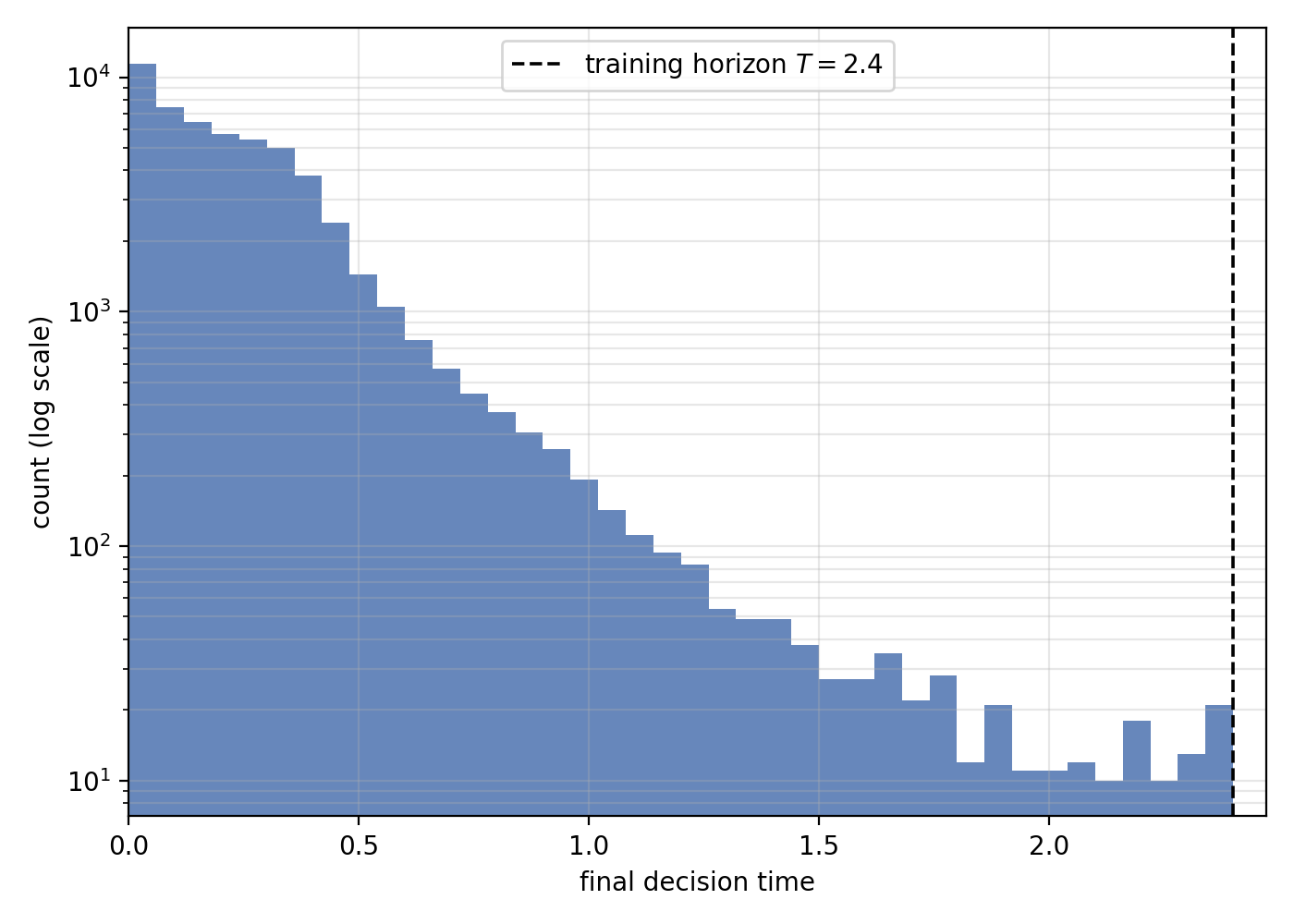}
        \caption{Training set}
        \label{fig:fmnist_decision_time_train}
    \end{subfigure}
    \hfill
    \begin{subfigure}{0.48\linewidth}
        \centering
        \includegraphics[width=\linewidth]{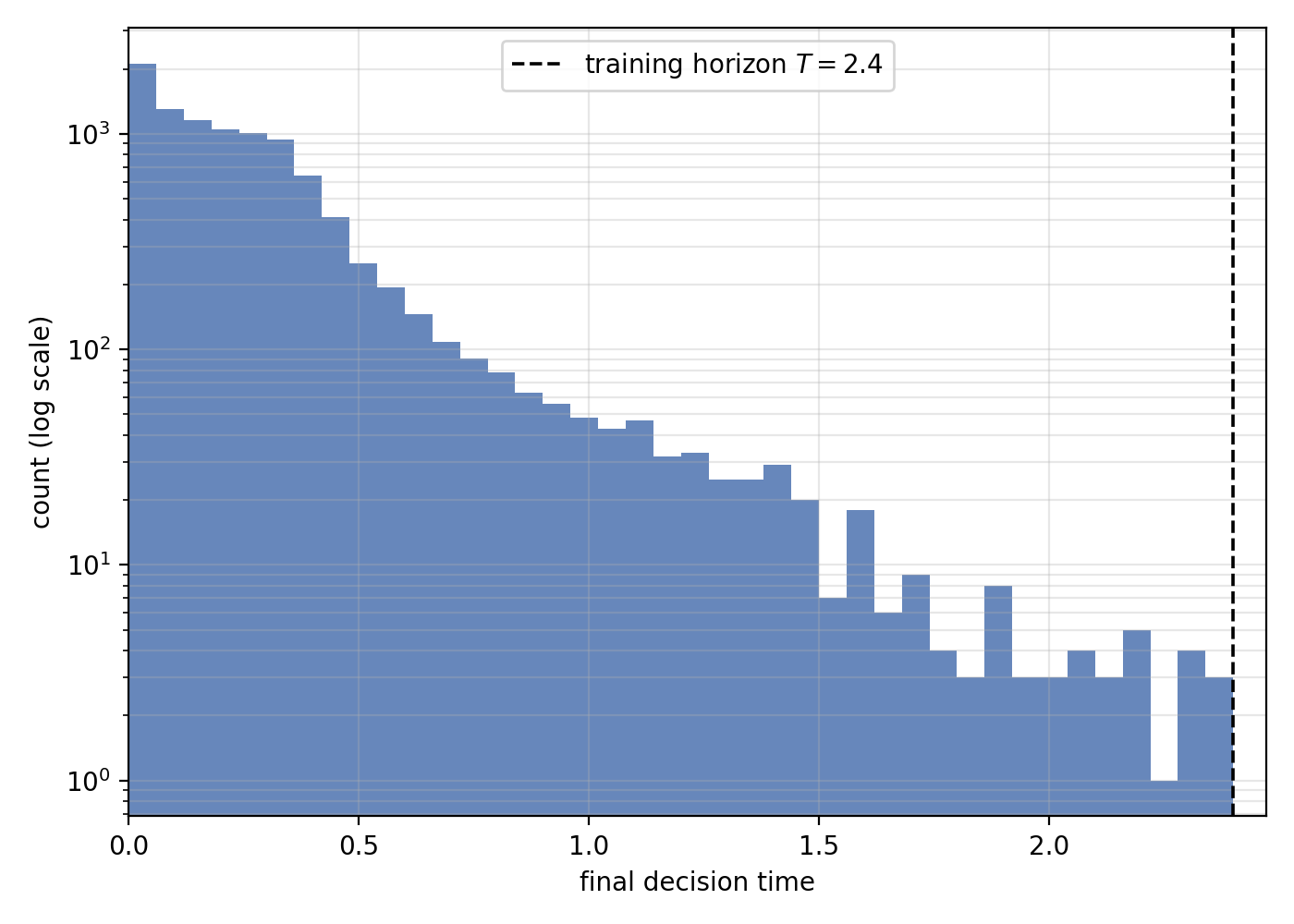}
        \caption{Test set}
        \label{fig:fmnist_decision_time_test}
    \end{subfigure}
    \caption{Distribution of the final decision times $\tau_i$ on Fashion-MNIST, for one of the five models, on the training set (a, $54{,}000$ images) and on the test set (b, $10{,}000$ images). Same definition as in Fig.~\ref{fig:decision_time_spirals}, the switches between two wrong classes included; the histogram has $40$ bins of width $2\Delta t=0.06$, and the inputs whose label never changes ($\tau_i=0$, about $13\,\%$) fall in the first one. The vertical axis is logarithmic, and the dashed line marks the training horizon $T=2.4$.}
    \label{fig:fmnist_decision_time}
\end{figure}

The decision times confirm that the labels are fixed early (Fig.~\ref{fig:fmnist_decision_time}): as for the spirals with one turn, their distribution decays approximately exponentially, with a median of about $T/12$. Only $2.3\,\%$ of the labels change after the shortened horizon $t_{\mathrm{best}}$ of Sec.~\ref{subsec:fmnist_bench}, in both directions, so that the accuracy is essentially unaffected (Tables~\ref{tab:fashionmnist_compare} and~\ref{tab:fashionmnist_tbest}). Finally, at $t=5T$ every test trajectory has reached one of the planted equilibria: unlike the planar model of the disk and the annulus (Sec.~\ref{subsec:trap_disk_2d}), no attractor other than the planted ones captures the data.

\clearpage
\bibliographystyle{plain}
\bibliography{references}

\end{document}